\documentclass[aip,reprint]{revtex4-1}
\usepackage{graphicx}
\usepackage{multirow}
\usepackage{amsmath}
\usepackage{xcolor}
\draft % marks overfull lines with a black rule on the right

\begin{document}

\title{Coupled cluster Green's function: Past, Present, and Future}
\author{Bo Peng}
\email{peng398@pnnl.gov} 
\address{Physical Sciences and Computational Division, Pacific Northwest National Laboratory, Richland, Washington 99354, United States of America}
\author{Nicholas P Bauman}
\address{Physical Sciences and Computational Division, Pacific Northwest National Laboratory, Richland, Washington 99354, United States of America}
\author{Sahil Gulania}
\address{Computational Science Division, Argonne National Laboratory, Lemont, IL 60439}      
\author{Karol Kowalski}
\address{Physical Sciences and Computational Division, Pacific Northwest National Laboratory, Richland, Washington 99354, United States of America}

\begin{abstract}
    Coupled cluster Green's function (CCGF) approach has drawn much attention in recent years for targeting the molecular and material electronic structure problems from a many-body perspective in a systematically improvable way. Here, we will present a brief review of the history of how the Green's function method evolved with the wavefunction, early and recent development of CCGF theory, and more recently scalable CCGF software development. We will highlight some of the recent applications of CCGF approach and propose some potential applications that would emerge in the near future.
\end{abstract}


\maketitle

\tableofcontents

\section{Introduction}

\subsection{Quantum Green's function method and its convolution with the wavefunction}

In his seminal work, ``An essay on the application of mathematical analysis to the theories of electricity and magnetism'' (1828)\cite{green2008essay}, George Green (1793$-$1841), an English miller, physicist, and (possibly) self-taught mathematician\cite{cannell2001Green}, introduced what has been known today as the Green's function, and developed a theory for solving the partial differential equations with general boundary conditions. This work was first self-published by Green himself and only circulated around family members and friends. Green's seminal work was rediscovered and recognized by Lord Kelvin, Sturm, Liouville, Dirichlet, Riemann, Neumann, Maxwell, and many other mathematicians and physicists\cite{Lindstrm2008OnTO,grattan1995why,jahnkehistory,nearing2010mathematical}. Since then, the Green's function method has been greatly developed and well established as a powerful mathematical tool for solving inhomogeneous differential equations and has been been widely used in the quantum many-body theory.

Formally, the Green’s function was developed as auxiliary functions for solving a linear differential equation with a Dirac delta inhomogeneous source and homogeneous boundary conditions. Loosely speaking, suppose we would like to solve a partial linear inhomogeneous differential equation 
\begin{align}
    \mathcal{L} p(x) = q(x)
\end{align}
with $\mathcal{L}$ being a linear differential operator, and $p(x)$ and $q(x)\neq0$ the solution and inhomogeneity source, respectively. Utilizing the Green's function $G(x,x')$, defined as the solution of similar linear differential equation with a delta inhomogeneity
\begin{align}
    \mathcal{L} G(x,x') = \delta(x-x') = \left\{
    \begin{array}{cc}
        \infty, & x = x'  \\
        0, & x \neq x'
    \end{array} \right. ,
\end{align}
a solution of $p(x)$ can then be formally found through
\begin{align}
    p(x) = \int G(x,x') q(x') {\rm d}x'.
\end{align}

Since it was born, the Green's function method had played a key role in solving the boundary-value problems in acoustics, hydrodynamics, thermodynamics, and electromagnetism during the second half of the 19th century\cite{nearing2010mathematical}. By the beginning of the 20th century, the Green's function technique had paved the foundation for the whole theory of the partial differential equation and the development of functional analysis. In particular, it was generalized to linear operator theory and applied to second-order linear differential equations. With the emergence of quantum mechanics during the same time, the Green's function manifested its power for dealing with the Schr\"{o}dinger equation\cite{schrodinger2003collected,amrein2005sturm} by converting the differential equation into integral operator problems\cite{Lindstrm2008OnTO}. In this context, the time-dependent Schr\"{o}dinger equation (TDSE) is viewed as a linear partial differential equation that is of first-order in time, 
\begin{align}
    \Bigg[ i\hbar\frac{\partial}{\partial \rm t} + \frac{\hbar^2}{2m}\nabla^2 \Bigg] \Psi(\textbf{r},\rm t) = V(\textbf{r},\rm t)\Psi(\textbf{r},\rm t) ,\label{se1}
\end{align}
where $\Psi(\textbf{r},\rm t)$ represents the time evolution of the wavefunction, $\frac{h^2}{2m}\nabla^2$ is the kinetic energy operator with the momentum operator $\textbf{p} = -i\hbar \nabla$, and $V(\textbf{r},\rm t)$ is the external potential term. To apply the Green's function method, one can first transform Eq. (\ref{se1}) to an inhomogeneous problem with Green's function being defined as its solution,
\begin{align}
    \Bigg[ i\hbar\frac{\partial}{\partial \rm t} + \frac{\hbar^2}{2m}\nabla^2 \Bigg] G(\textbf{r},\rm t; \textbf{r}',{\rm t}') = \delta(\textbf{r}-\textbf{r}')\delta({\rm t}-{\rm t}'). \label{se2}
\end{align}
The inhomogeneous solution $\Psi(\textbf{r},\rm t)$ may then be obtained through
\begin{align}
    \Psi(\textbf{r},\rm t) = \int G(\textbf{r},\rm t; \textbf{r}',{\rm t}')\Psi(\textbf{r}',{\rm t}'){\rm d}\textbf{r}'. \label{wfn1}
\end{align}
Eq. (\ref{wfn1}) can also be interpreted as the time evolution of the wavefunction from a given time and position $(\textbf{r}',{\rm t}')$ to another time and position $(\textbf{r},{\rm t})$. 

Note that the wavefunction $\Psi(\textbf{r},\rm t)$ in the Schr\"{o}dinger picture can be rewritten as time propagation of state vectors in the position representation, i.e.
\begin{align}
    \Psi(\textbf{r},\rm t) 
    &= \langle \textbf{r} | \Psi(\rm t) \rangle = \langle \textbf{r} | U ({\rm t}, {\rm t}')\Psi({\rm t}') \rangle \notag \\
    &= \int \langle \textbf{r} | U ({\rm t}, {\rm t}') | \textbf{r}'\rangle \langle \textbf{r}' | \Psi({\rm t}')\rangle {\rm d} \textbf{r}' \notag \\
    &= \int \langle \textbf{r} | U ({\rm t}, {\rm t}') | \textbf{r}'\rangle \Psi(\textbf{r}',{\rm t}') {\rm d} \textbf{r}' \label{wfn2} ,
\end{align}
where the closure relation $\int | \textbf{r}'\rangle \langle \textbf{r}' | {\rm d}\textbf{r}' = \mathbf{1}$ is applied, and the time evolution operation $U ({\rm t}, {\rm t}') = e^{-\frac{i}{\hbar}H({\rm t}-{\rm t}')}$ in the Schr\"{o}dinger picture evolves the wavefunction $\Psi(\textbf{r}',{\rm t}')$ to $\Psi(\textbf{r},{\rm t})$ in infinitesimal time invervals.
Comparing Eq. (\ref{wfn2}) with Eq. (\ref{wfn1}), we can then write
\begin{align}
    G(\textbf{r},\rm t; \textbf{r}',{\rm t}') = \langle \textbf{r} | U ({\rm t}, {\rm t}') | \textbf{r}'\rangle = \langle \textbf{r},{\rm t} | \textbf{r}',{\rm t}'\rangle, \label{gf1}
\end{align}
which associates the Green's function to the probability amplitude of finding the particle in a state $| \textbf{r},{\rm t}\rangle$ given its initial state $| \textbf{r}',{\rm t}'\rangle$\cite{dirac1933Lagrangian}.
Since the Hamiltonian can be represented as $H = \sum_n E_n | n \rangle \langle n |$ with $(E_n, |n\rangle)$'s being its eigenpairs, and the eigenstates, $|n\rangle$'s, form a complete set, i.e. $\sum_n | n \rangle \langle n | = \mathbf{1}$, we can further rewrite Eq. (\ref{gf1}) as
\begin{align}
    G(\textbf{r},\rm t; \textbf{r}',{\rm t}') 
    &= \sum_n \langle \textbf{r} | e^{-\frac{i}{\hbar}H({\rm t}-{\rm t}')} | n \rangle \langle n | \textbf{r}'\rangle \notag \\
    &= \sum_n \langle \textbf{r} | n \rangle \langle n | \textbf{r}'\rangle e^{-\frac{i}{\hbar}E_n({\rm t}-{\rm t}')} \notag \\
    &= \sum_n \psi_n(\textbf{r}) \psi_n^\ast (\textbf{r}') e^{-\frac{i}{\hbar}E_n({\rm t}-{\rm t}')}, \label{gfxn20}
\end{align}
whose Fourier transform then gives the spectral expansion
\begin{align}
    G(\textbf{r}, \textbf{r}';E) = \sum_n i \frac{\psi_n(\textbf{r}) \psi_n^\ast (\textbf{r}')}{E - E_n} \label{gfxn3}.
\end{align}
Here $\psi_n(\textbf{r})$ defines the projection $\langle \textbf{r} | n \rangle$. 

In the mid 20th century, the quantum Green’s functions were further developed as being introduced as quantum propagators in the quantum field theory by Feynman and Schwinger\cite{feynman1949theory,feynman1958Space,schwinger1959Theory} for the complete development of a path-integral formulation that interprets the wavefunction, i.e. Eq. (\ref{wfn1}), as the sum of the probabilities of the particle taking different quantum-mechanically possible trajectories. In particular, Feynman generalized the Green's function $G(\textbf{r},\rm t; \textbf{r}',{\rm t}')$ as a new quantum field propagator by accounting for the forward and backward propagation in space-time and causal time orderings. Martin and Schwinger also realized the importance of Green's function in quantum field theory and applied many-body Green's function in the condensed-matter physics\cite{schwinger1959Theory} to evaluate particle currents and spectral amplitudes. Furthermore, thermodynamic many-particle Green’s function using a grand-canonical ensemble average was also developed during this time by Kadanoff and Baym\cite{baym1961conservation}.

Since then, the quantum many-body Green’s function that connects different positions and times has stated to manifest its power in particle scattering, non-equilibrium and finite-temperature physics, quantum transport, and many other fields. In these studies, often time the solution of the problem is obtained for an effective single-particle wavefunction which is only exact for non-interacting system, and becomes approximate in interacting system. For example, the Landauer--B\"{u}ttiker method\cite{LB1,LB2,LB3} applied for the study of coherent quantum transport is usually not enough to describe electron-electron and electron-vibron interactions. When a single-particle picture no longer holds, explicit consideration of the wavefunction in the full many-body Hilbert space becomes necessary. 

\subsection{Coupled cluster Green's function, early work and recent development}

Among the well-developed explicit wavefunction approaches, the coupled cluster (CC) approach\cite{cizek66_4256,cizek1969,cizek1971,paldus72_50} is one of the most successful and efficient approaches that handle the many-body effect in a systematically improvable way. The marriage of the coupled cluster approach with the Green's function method, or namely the coupled cluster Green's function (CCGF) approach, dates back to the early 1990s when Nooijen and Snijders first showed in their seminal works\cite{nooijen92_55,  nooijen93_15, nooijen95_1681} that CCGF can be directly computed rather than being obtained through solving Dyson's equation with approximate self-energy. In the CCGF framework, the systematically improvable CC parametrization of the correlated wavefunction helps improve the computation of the Green's function in a systematically improvable manner. Also, due to the connection between the CCGF approach and the equation-of-motion coupled cluster (EOM-CC) approach, the pole positions in the CCGF structure representing ionization potentials or electron affinities can be exactly reproduced by the eigenvalues of the similarity transformed Hamiltonian in normal product form in ($N\pm1$)-particle Hilbert space. However, due to the high-order polynomial scaling of the CCGF approach with respect to the system size (e.g. for CCGF with singles and doubles, i.e. CCSDGF, the scaling is $\mathcal{O}(n^6)$ with $n$ being the number of basis functions representing the system size. For CCSDTGF, i.e. CCGF with singles, doubles, and triples, the scaling becomes $\mathcal{O}(n^8)$) and limitation of computing software and hardware,
early CCGF practice was subject to only resolving limited ionized states of small molecules described by small basis sets. 

During the last 30 years, one has witnessed the rapid development of high-performance computing software and hardware that boosts size limit and time scale in quantum simulations. In recent years, CCGF starts to regain attention in molecular and material quantum chemical calculations. For example, for quantum systems described by $<$500 basis functions, one can now routinely employ CCSDGF approach to compute the corresponding many-body electronic structures, and the spectral function calculations have been reported for systems ranging from uniform electron gas\cite{chan16_235139}, light atoms\cite{matsushita18_034106}, heavy metal atoms\cite{matsushita18_224103}, and simple 1-D periodic systems \cite{matsushita18_204109}, to a bunch of small and medium-size molecular systems\cite{kowalski18_4335,berkelbach18_4224}. More recently, the CCGF impurity solver has been reported for computing the electronic structure of complex materials in the embedding computing framework\cite{zhu2019_115154,zgid19_6010}. However, large CCGF calculations with more than 1000 basis functions are still challenging today, and one would need to introduce further approximations that might compromise the accuracy or employ supercomputing facility to perform ``hero" runs. For example, with the aid of Oakridge leadership computing facility\cite{olcfsummit} and deeply optimized numerical library specializing CCGF calculations\cite{peng2021GFCCLib}, Peng et al. reported the CCSDGF calculations for deciphering the valence band electronic structures of C60 molecules and a series of large DNA fragments described with as many as $\sim$1200 basis functions\cite{peng20_011101}.

In the following, we will first review the theoretical foundation of the CCGF approach. After that, we will give a discussion about the numerical approach and the development of relevant software. Then we will elaborate on some recent applications employing the CCGF approach. At last, we will conclude by proposing and outlooking some potential theoretical developments and applications in this area in the future.

\section{Theoretical foundation and Numerical approaches}

\subsection{CCGF theory}

In the following, we give a brief introduction of the CCGF theory. For a detailed review of the GFCC method employed in this work, we refer the readers to Refs. \onlinecite{nooijen92_55, nooijen93_15, nooijen95_1681,meissner93_67,kowalski14_094102, kowalski16_144101,kowalski16_062512, kowalski18_561,kowalski18_4335, kowalski18_214102}. Following Eq. (\ref{gfxn20}), the analytical time-dependent Green's function of an $N$-electron system that is governed by Hamiltonian $H$ at the frequency $\omega$ can be expressed as
\begin{eqnarray}
G_{pq}(t) &=& \langle \Psi | a_p e^{+\text{i}2\pi(H-E_0)t} a_q^\dagger | \Psi \rangle \notag \\
&+&\langle \Psi | a_q^\dagger e^{-\text{i}2\pi(H-E_0)t} a_p | \Psi \rangle . \label{gfxn_t}
\end{eqnarray} 
The corresponding Fourier transform then gives rise to the CCGF in the frequency domain
\begin{eqnarray}
G_{pq}(\omega) =
\langle \Psi | a_p (\omega - ( H - E_0 ) + i \eta)^{-1} a_q^\dagger | \Psi \rangle \notag \\
+ \langle \Psi | a_q^\dagger (\omega + ( H - E_0 ) - i \eta)^{-1} a_p | \Psi \rangle.
\label{gfxn_omega}
\end{eqnarray}
Here, the $a_p$ ($a_p^\dagger$) operator is the annihilation (creation) operator for electron in the $p$-th spin-orbital (we use $p,q,r,s,\ldots$ for the general spin-orbital indices, $i,j,k,l,\ldots$ for the occupied spin-orbital indices, and $a,b,c,d,\ldots$ for the virtual  spin-orbital indices), $E_0$ is the ground-state energy, and $\eta$ is the broadening factor introduced numerically to provide the width of the computed spectral bands. The normalized ground-state wavefunction of the system, $| \Psi \rangle$, is formulated through CC biorthogonal parametrization on the reference state $|\Phi\rangle$, i.e. 
\begin{align}
    | \Psi \rangle = e^T | \Phi \rangle,~~~~\langle \Psi | = \langle \Phi | (1+\Lambda) e^{-T},
\end{align}
where the cluster operator $T$ and the de-excitation operator $\Lambda$ are obtained from solving the conventional CC equations. Eqs. (\ref{gfxn_t}) and (\ref{gfxn_omega}) can be further transformed to a compact form
\begin{eqnarray}
G_{pq}(t) = 
\langle\Phi|(1+\Lambda) \overline{a_p} e^{+\text{i}2\pi \bar{H}_N t} \overline{a_q^{\dagger}} |\Phi\rangle \notag \\
+ \langle\Phi|(1+\Lambda) \overline{a_q^{\dagger}} e^{-\text{i}2\pi \bar{H}_N t} \overline{a_p} |\Phi\rangle
\label{gfxn0}
\end{eqnarray}
\begin{eqnarray}
G_{pq}(\omega) = 
\langle\Phi|(1+\Lambda) \overline{a_p} (\omega-\bar{H}_N + \text{i} \eta)^{-1} \overline{a_q^{\dagger}} |\Phi\rangle \notag \\
+ \langle\Phi|(1+\Lambda) \overline{a_q^{\dagger}} (\omega+\bar{H}_N - \text{i} \eta)^{-1} \overline{a_p} |\Phi\rangle,
\label{gfxn1}
\end{eqnarray}
with the normal product form of similarity transformed Hamiltonian $\bar{H}_N$ being defined as $\bar{H} - E_0$. Here, the similarity transformed operators $\bar{A}$ ($A = H, a_p, a_q^{\dagger}$) are defined as $\bar{A} = e^{-T} A ~e^{T}$. The self-energy can be obtained directly through Dyson's equation, 
\begin{align}
    {\bf \Sigma} = {\bf G}_0^{-1} - {\bf G}^{-1}.
\end{align}
Here ${\bf G}_0$ is the reference Green's function chosen to be free-particle Helmholtz equation.

\subsection{Numerical approaches for computing CCGF}

The development of the numerical approaches following Eqs. (\ref{gfxn_t}) and (\ref{gfxn0}) for computing time-dependent CCGF is still at its early stage, and has only been studied for model systems\cite{keen2021hybrid} and small molecular systems\cite{rehr2020eom,vila2020real}. For the latter, the reported studies are focused on propagating the state in the ($N-1$)-electron space within the cumulant approximation at the CC level to resolve the real-time evolution of the core ionized state, in particular for the simulation of the x-ray absorption spectra (XAS) from a deep core level. In the following, we will mainly focus on the review of the intensively developed numerical approaches for the CCGF calculations in the frequency space in recent years. 

In the frequency space, to compute the $\omega$-dependent CCGF matrix, in a straightforward manner, one can diagonalize the non-Hermitian similarity transformed Hamiltonian $\bar{H}$ in the ($N \pm 1$)-particle space to construct the sum-over-states representation
\begin{align}
    G_{pq}(\omega) = \sum_{\mu=1}^{M^{(N+1)}} \frac{\langle \Psi | \overline{a_p} | R_{\mu}^{(N+1)} \rangle \langle L_{\mu}^{(N+1)} | \overline{a_q^{\dagger}} | \Psi \rangle }{\omega - (E_{\mu}^{(N+1)} - E_0) + {\rm i}\eta} \notag \\
    + \sum_{\mu=1}^{M^{(N-1)}} \frac{\langle \Phi | \overline{a_q^{\dagger}} | R_{\mu}^{(N-1)} \rangle \langle L_{\mu}^{(N-1)} | \overline{a_p} | \Phi \rangle }{\omega - (E_0 - E_{\mu}^{(N-1)}) - {\rm i}\eta}. \label{sos}
\end{align}
Here $R_{\mu}^{(N\pm 1)}$'s and $L_{\mu}^{(N\pm 1)}$'s are right and left eigenvectors of $\bar{H}$ in the ($N\pm 1$)-electron space with $E_{\mu}^{(N\pm 1)}$'s the corresponding eigenvalues and $M^{(N\pm 1)}$ the total number of eigenpairs.
Therefore, if based on Eq. (\ref{sos}), direct computing of the CCGF matrix would require the diagonalization of the similarity transformed Hamiltonian matrix to obtain the non-negligible excited state eigenvalues and eigenvectors. Unfortunately, as the dimension of the Hamiltonian matrix grows polynomially with the number of orbitals used to describe the system, solving such an eigen-problem, if the Hamiltonian is a large dense matrix (e.g. dimension $10^{10}$), would become challenging. In this scenario, direct diagonalization via, for example, LAPACK and ScaLAPACK quickly become infeasible, and one may need to switch to utilize standard matrix-free iterative Krylov subspace methods such as Arnoldi's methods and variants\cite{arno1951,bade2000} to obtain the eigenpairs in the extreme ends of the spectrum of the Hamiltonian. Obviously, there are two clear disadvantages in this sum-over-states approach, (1) the choice of states to be included in Eq. (\ref{sos}) may be strongly $\omega$-dependent, and there are no clear criteria for how many states would need to be included in Eq. (\ref{sos}), which jeopardizes its practical applications. (2) Due to the polynomial scaling of the CC calculations and the large dimension of the Hamiltonian matrix, only a limited number of states can be obtained with the help of iterative methods, and if the target states are embedded deep in the spectral interior of the Hamiltonian, the convergence performance of the standard Arnoldi's method would significantly deteriorate. For the latter, to reduce scaling and improve performance, one might need to employ spectral transformations\cite{bade2000} or introduce other physics-inspired approximations. For example, early attempts perturbatively treated the sub-blocks of the similarity transformed Hamiltonian, and achieved an $\mathcal{O}(n^5)$ scaling by treating the double-double block as diagonal (or the zeroth order) and truncating other blocks at the second order.\cite{stanton95_1064} 

Another typical approximation is core-valence separation (CVS)\cite{cederbaum80_206} which has been routinely applied in the CC methods and algebraic-diagrammatic construction (ADC) methods\cite{norman18_7208}. In the approximation, the coupling between core- and valence-excited states is neglected to reduce the dimension of the effective Hamiltonian and accelerate the convergence. It has been reported that the typical error brought by the CVS approximation in the calculations of the K-shell ionization spectra of small and medium-size molecules\cite{schirmer87_6789, dreuw14_4583, trofimov00_483} would be between 0.4 to 1.0 eV. Nevertheless, for smaller errors and higher resolution of the spectra, the CVS may not be sufficient and other robust methods would be required. In the context of the relevant equation-of-motion (EOM) and linear response (LR) CC calculations, recent years have witnessed the development of some variants of the Krylov subspace methods along this line$-$asymmetric Lanczos-chain-driven subspace algorithm\cite{coriani12_1616}, energy-specific Davidson algorithm\cite{peng15_4146}, and Generalized Preconditioned Locally Harmonic Residual (GPLHR) method\cite{krylov15_273} to name a few. Remarkably, to serve the reduced scaling purpose in the coupled cluster calculations, local descriptions of the correlation wavefunction, for example, the pair natural orbitals (PNOs) \cite{meyer73_1017,edmiston66_1833,edmiston68_192,ahlrichs75_275} or its modern version domain-based local pair natural orbitals (DLPNO) \cite{neese13_034106,neese16_024109} can be further used. There have been reports using these techniques to reduce the size of the space for the diagonalization of the similarity transformed Hamiltonian. \cite{neese16_034102,neese18_244101,neese19_164123}.

\begin{figure}[h!]
    \includegraphics[width=\linewidth]{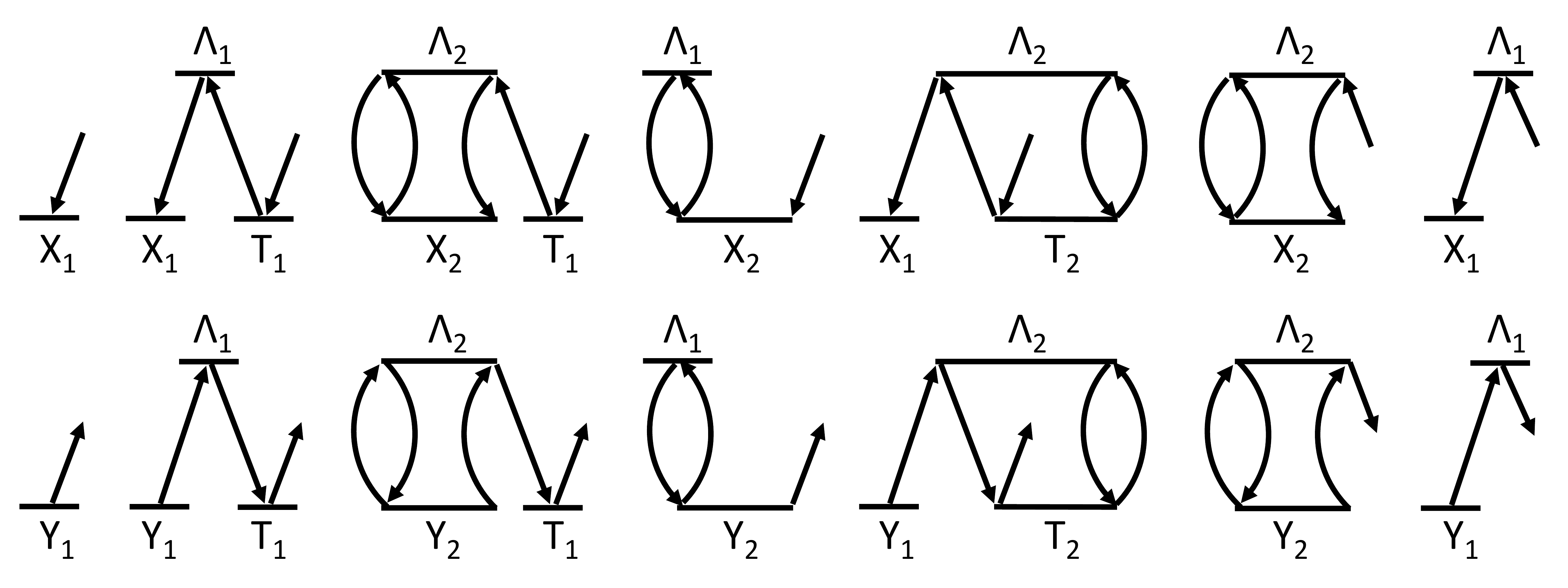}
    \caption{All the connected diagrams contributing to the CCSDGF matrix.}
    \label{CCSDGF_diag}
\end{figure}

Beside computing the CCGF matrix through an eigen-solving process, the analytical CCGF form in Eq. (\ref{gfxn1}) also allows computing CCGF matrix in the frequency domain through linear solvers\cite{kowalski14_094102, kowalski16_144101,kowalski16_062512, kowalski18_561,kowalski18_4335, kowalski18_214102}. 
In this approach, one need to first obtain the auxiliary many-body solutions $X_p(\omega)$ and $Y_q(\omega)$ through solving the following linear systems in the ($N\pm 1$)-particle space
\begin{eqnarray}
(\omega-\bar{H}_N + \text{i} \eta )Y_q(\omega)|\Phi\rangle &=& \overline{a_q^\dagger} |\Phi\rangle, \notag \\
(\omega+\bar{H}_N - \text{i} \eta )X_p(\omega)|\Phi\rangle &=& \overline{a_p} |\Phi\rangle. \label{eq:xplin} 
\end{eqnarray}
Then the CCGF matrix can be computed through contraction
\begin{eqnarray}
G_{pq}(\omega) &=& \langle\Phi|(1+\Lambda) \overline{a_p} Y_q(\omega) |\Phi\rangle \notag \\
&+& \langle\Phi|(1+\Lambda) \overline{a_q^{\dagger}} X_p(\omega) |\Phi\rangle.
\label{gfxn2}
\end{eqnarray}
Here, the structures of $\omega$-dependent many-body operators $X_p(\omega)$ and $Y_q(\omega)$ can be expressed as
\begin{eqnarray}
Y_q(\omega) 
&=& Y_{q,1}(\omega)+Y_{q,2}(\omega) + \ldots \notag \\
&=& \sum_{a} y_a(\omega)_q  a_a^\dagger  + \sum_{i,a<b} y^{i}_{ab}(\omega)_q a_a^{\dagger} a_b^{\dagger} a_i +\ldots , \label{yq} \\
X_p(\omega) 
&=& X_{p,1}(\omega)+X_{p,2}(\omega) + \ldots \notag \\
&=& \sum_{i} x^i(\omega)_p  a_i  + \sum_{i<j,a} x^{ij}_a(\omega)_p a_a^{\dagger} a_j a_i +\ldots , \label{xp} 
\end{eqnarray}
which similar to the cluster amplitudes $T$ and $\Lambda$ can be truncated at different many-body levels. At the two-body level, the obtained approximate formulation of Eq. (\ref{gfxn2}) can be writen as
\begin{eqnarray}
G_{pq} &=&  
\langle\Phi|(1+\Lambda_1+\Lambda_2) \overline{a_p} (Y_{q,1}+Y_{q,2}) |\Phi\rangle  \notag \\
&+& \langle\Phi|(1+\Lambda_1+\Lambda_2) \overline{a_q^{\dagger}} (X_{p,1}+X_{p,2}) |\Phi\rangle \label{gfxn3},
\end{eqnarray}
which is the so-called CCSDGF approximation (CCGF with singles and doubles). Fig. \ref{CCSDGF_diag} exhibits all the connected diagrams contributing to the CCSDGF matrix.

Comparing the eigen-solving and linear-solving approaches towards computing CCGF matrix, we can see that (1) the eigen-solving process needs to resolve all the necessary states or the necessary bases (e.g. Lanczos bases) for reproducing all the states to suppress the error due to the missing states, and (2) the linear-solving approach bypasses the ambiguity of choosing the $\omega$-dependent states to be included in Eq. (\ref{sos}) while requires to sweep a range of frequency to resolve the pole positions. Therefore, during the linear-solving process the number of frequencies (or the number of the sampled $k$ points when dealing with a periodic system) will impose a larger overhead on the cost of computing CCGF matrix if a much higher resolution is required. To solve this problem, recent computation for the CCGF of periodic systems showed that the self-energy-mediated interpolation utilizing Wannier orbitals to compute Green's function via Dyson equation can be used to accommodate a large number of sampled $k$ points.\cite{yuichiro18}
More recently, a more general interpolation technique has been reported in the CCGF calculations of small and medium-size molecular systems utilizing the so-called  model-order-reduction (MOR) technique\cite{peng19_3185}. Essentially, MOR helps projects the target linear problem to Krylov subspace such that the solution of the original linear system over a broad frequency regime can be approximated through a more easily solvable model linear system over the same frequency regime. For the CCGF-associated linear problem, the MOR projection can be expressed as
\begin{eqnarray}
\left\{ \begin{array}{ccl}
(\omega - \text{i} \eta + \hat{\bar{\textbf{H}}}_N) \hat{\textbf{X}} & = & \hat{\textbf{b}_1} \\
(\omega + \text{i} \eta - \hat{\bar{\textbf{H}}}_N) \hat{\textbf{Y}} & = & \hat{\textbf{b}_2}
\end{array}\right. , \label{mor}
\end{eqnarray}
and
\begin{eqnarray}
\hat{\mathbf{G}} & = & \hat{\mathbf{c}_1}^{\text{T}} \hat{\textbf{X}} + \hat{\mathbf{c}_2}^{\text{T}} \hat{\textbf{Y}},
\label{model}
\end{eqnarray}
where an orthonormal subspace $\mathbf{S}=\{\mathbf{v}_1, \mathbf{v}_2, \ldots, \mathbf{v}_m\}$ (the rank $m$ is much smaller than the dimension of the Hamiltonian) is construct to project $\hat{\bar{\textbf{H}}}_N = \textbf{S}^{\text{T}} \bar{\textbf{H}}_N \textbf{S}$, $\hat{\textbf{X}} = \textbf{S}^{\text{T}} \textbf{X}$, $\hat{\textbf{Y}} = \textbf{S}^{\text{T}} \textbf{Y}$, $\hat{\textbf{b}_i} = \textbf{S}^{\text{T}} \textbf{b}_i$, and $\hat{\mathbf{c}_i}^{\text{T}} = \mathbf{c}_i^{\text{T}} \textbf{S}$ ($i = 1,2$) with the columns of $\textbf{b}_1$ and $\textbf{b}_2$ corresponding to $\overline{a_p} |\Phi\rangle$ or $\overline{a_q^\dagger} |\Phi\rangle$, and the columns of $\textbf{c}_1$ and $\textbf{c}_2$ corresponding to $\langle \Phi | (1+\Lambda) \overline{a^\dagger_q}$ or $\langle \Phi | (1+\Lambda) \overline{a_p}$, respectively. 

The overall cost of performing CCGF simulations comes from several parts. First of all, the conventional Hartree-Fock (HF) and CC calculations (this could also include $\Lambda$-CC calculations) need to be performed to prepare $|\Psi\rangle$. The conventional HF calculation scales as $\mathcal{O}(n^3)$. The conventional ($\Lambda$-)CCSD calculation  scales as $\mathcal{O}(n^6)$. Secondly, the linear-solving process in the $\omega$-dependent CCGF calculation needs to be performed for all the spin-orbitals and frequencies. For a given frequency, one CCSDGF calculation scales as $\mathcal{O}(n^6)$. In addition, there is a prefactor, the number of frequencies, applying on the top of the cost. If applying MOR, the prefactor can be significantly reduced, with the trade-off being the orthonormalization of the subspace, subspace projection, and solving Eq. (\ref{mor}). Typically, a Gram-Schmidt (GS) orthogonalization would scale as $\mathcal{O}(m^2n^3)$, the projection scales as $\mathcal{O}(mn^5)$, and solving the projected linear system scales as $\mathcal{O}(N_{\omega}m^3)$ with $N_{\omega}$ the total number of frequencies.

From the above analysis, it is obvious that the practical CCGF calculation is mainly subject to (a) intensive compute/contraction and communication of multi-dimension tensors and (b) performance of the iterative solver. Remarkably, both (a) and (b) are shared between CC and CCGF calculations, and can be utilized to ease the implementation. For example, the direct inverse of iterative subspace (DIIS) solver\cite{pulay80_393,pulay82_556} is usually employed in the non-linear ground-state CC calculations. For the CCGF calculations, it has been reported that DIIS solver that is also employed to solver CCGF linear systems exhibited faster convergence than the Lanczos iterative method for computing the valence ionization potentials of CO and N$_2$ molecules\cite{kowalski18_4335,kowalski18_214102}. Other solvers such as biconjugate gradient (BiCG)\cite{matsushita18_034106} and generalized minimal residual (GMRes) method\cite{peng20_011101} have also been reported in the CCGF calculations. In particular, GMRes is the only Krylov-based solver for treating general matrices (note that the similarity transformed hamiltonian in GFCC is non-symmetric) and, in principle, able to converge to the exact solution after at most $\mathcal{O}(O^2V)$ steps. However, in comparison with the DIIS whose cost is constant about $\mathcal{O}(O^2V)$, the cost of GMRes grows as $\mathcal{O}(k^2O^2V)$ with $k$ being the iteration number.

%%%%%%%%%%%%%%%%%%%%%%%%%%
\subsection{The state-of-the-art implementation and software development towards scalable simulations}

\begin{figure}[h!]
    \includegraphics[width=\linewidth]{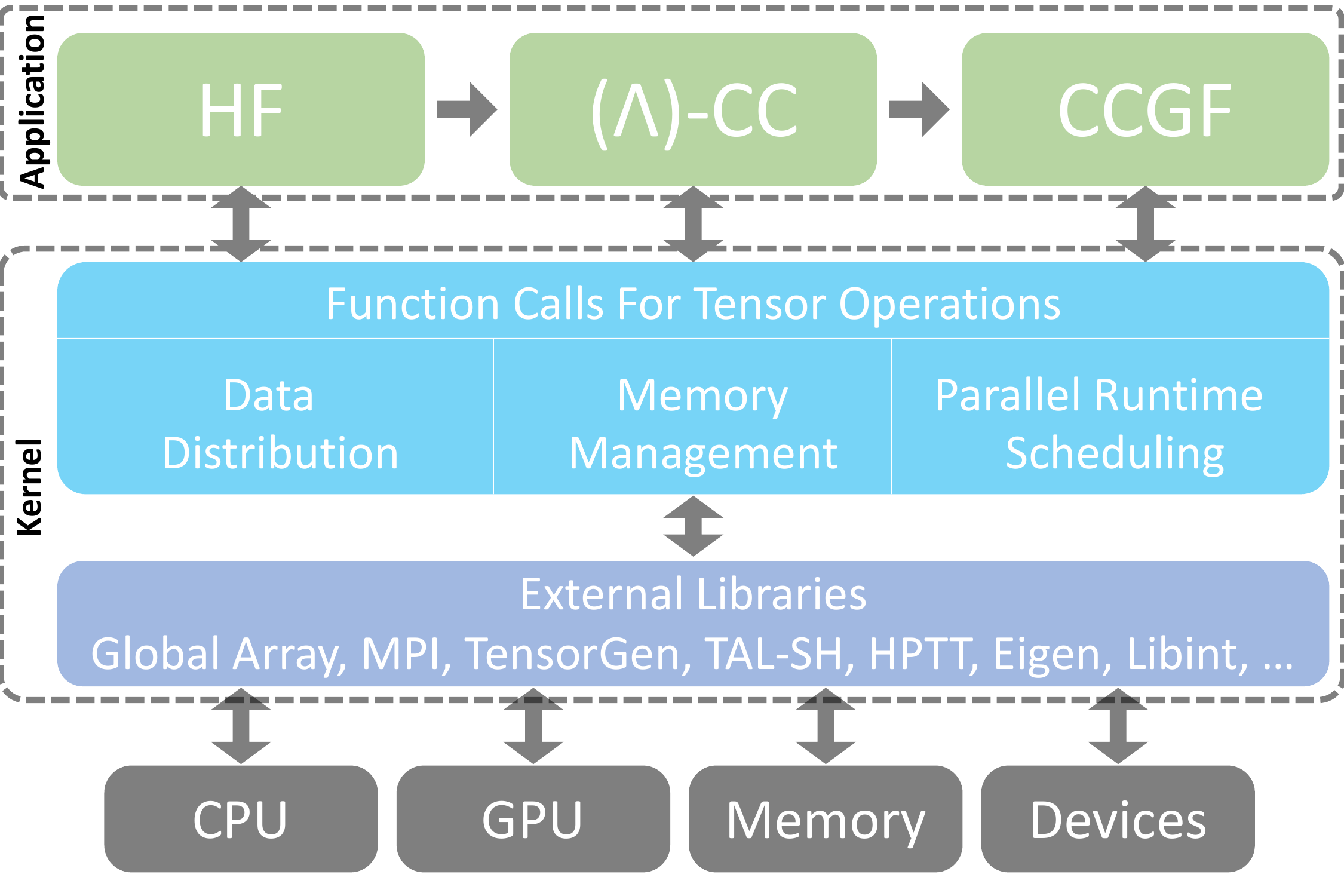}
    \caption{Infrastructure of GFCCLib\cite{peng2021GFCCLib}, a library package for performing scalable CCGF calculations.}
    \label{CCSDGF_module}
\end{figure}

In the frequency space, the algebraic structure of the $\omega$-dependent CCGF equations intuitively enables  scalable implementation$-$parallel execution through frequencies and/or spin-orbitals. However, a rudimentary implementation based on the intuitive parallelism would quickly encounter bottlenecks, in particular, the ones associated with expensive high dimensional tensor contractions, expensive inter-processor communication, and severe load imbalance, which prohibit the development of highly scalable CCGF library and its routine application in large scale simulations. Based on our previous experience on designing and developing computational chemistry packages to work efficiently on massively parallel processing  supercomputers (e.g NWChem\cite{valiev2010nwchem} and more recently NWChemEx\cite{nwchemex}), properly addressing these bottlenecks requires developments of new computational strategy and solvers, and the systematic software engineering of an up-to-date inclusive computing infrastructure that efficiently utilizes the highly scalable computing resource. Especially, the efficient implementations of CCGF utilizing presently available supercomputers need to focus on the specialized form of tensor libraries to provide tools  for distributing multidimensional tensors across the network, performing tensor contractions in parallel, and  utilizing computational resources offered by existing GPU architectures. Take the high dimensional tensor contractions for an example, the most expensive tensor contraction in the CCSDGF approach can be expressed as 
\begin{eqnarray}
R(i,j,a,b) &+=& \sum_{c,d} ~T(i,j,c,d)\times V(a,b,c,d) \label{cont1}
\end{eqnarray}
where $V$ and $T$ are four-index two-electron integral tensor and CC double amplitude tensor, respectively, and $R$ is an four-index intermediate tensor. The multidimensional tensor contraction shown in Eq. (\ref{cont1}) is both computation and communication intensive. In the early developments of parallel tensor algebra systems for dealing with this type of tensor contraction\cite{hirata039887, hirata2006symbolic, deumens2011software, deumens2011super, solomonik2013cyclops, solomonik2014massively, calvin2015scalable, peng2019coupled},
progress has been made towards automated code generators, memory reduction, and the real-space many-body calculations. 

Early this year, we released a new library, named GFCCLib\cite{peng2021GFCCLib}, for routinely performing CCGF calculations in scalable computing resources. Fig. \ref{CCSDGF_module} exhibits the infrastructure of GFCCLib. Basically, there are two layers inside the GFCCLib: (1) At the application layer, there are essentially three major modules to perform Hartree-Fock (HF), ground-state CC, and CCGF calculations, respectively; (2) The application layer routinely calls the functions provided by the kernel layer for performing tensor related operations. The tensor data distribution, memory management, and parallel runtime scheduling are dealt with inside the kernel through invoking external vendor libraries. Specifically, the kernel, which can also be viewed as an independent tensor algebra library for many-body methods\cite{mutlu2019toward}, is implemented using Global Arrays (GA)\cite{ga1994,ga1995,ga1996} and message passing interface (MPI) targeting high-performance parallelization on distributed memory platforms. Here, the kernel API directly takes the distributed block-sparse tensors and decomposes the contractions into a set of dependent operations that are then passed to a backend for scheduling and execution. High performance is obtained in the backend by focusing upon a small number of operations that are extensively optimized by the vendor libraries or by code generation plus auto-tuning. Also, inside the kernel, the execution of tensor operations involves multi-granular dependence analysis and task-based execution. The dependence analysis splits the tensor operations into independent execution levels, and operations at the same level can be executed concurrently with synchronizations between levels. Note that this allows multiple operations to be executed at the same time, exposing more parallelism and improving processor utilization. Furthermore, each operation is partitioned into tasks that can be scheduled to compute a portion of the operation on any compute unit. To further speed up the tensor operation, a GPU execution and launch scheme (with TALSH \cite{TALSH} as the underlying GPU tensor algebra engine) is also developed inside the kernel to make use of the GPU acceleration for localized summation loops and enable data transfer and compute overlap between successive summation loop iterations.

\section{Recent applications}

\begin{figure}[h!]
    \includegraphics[width=\linewidth]{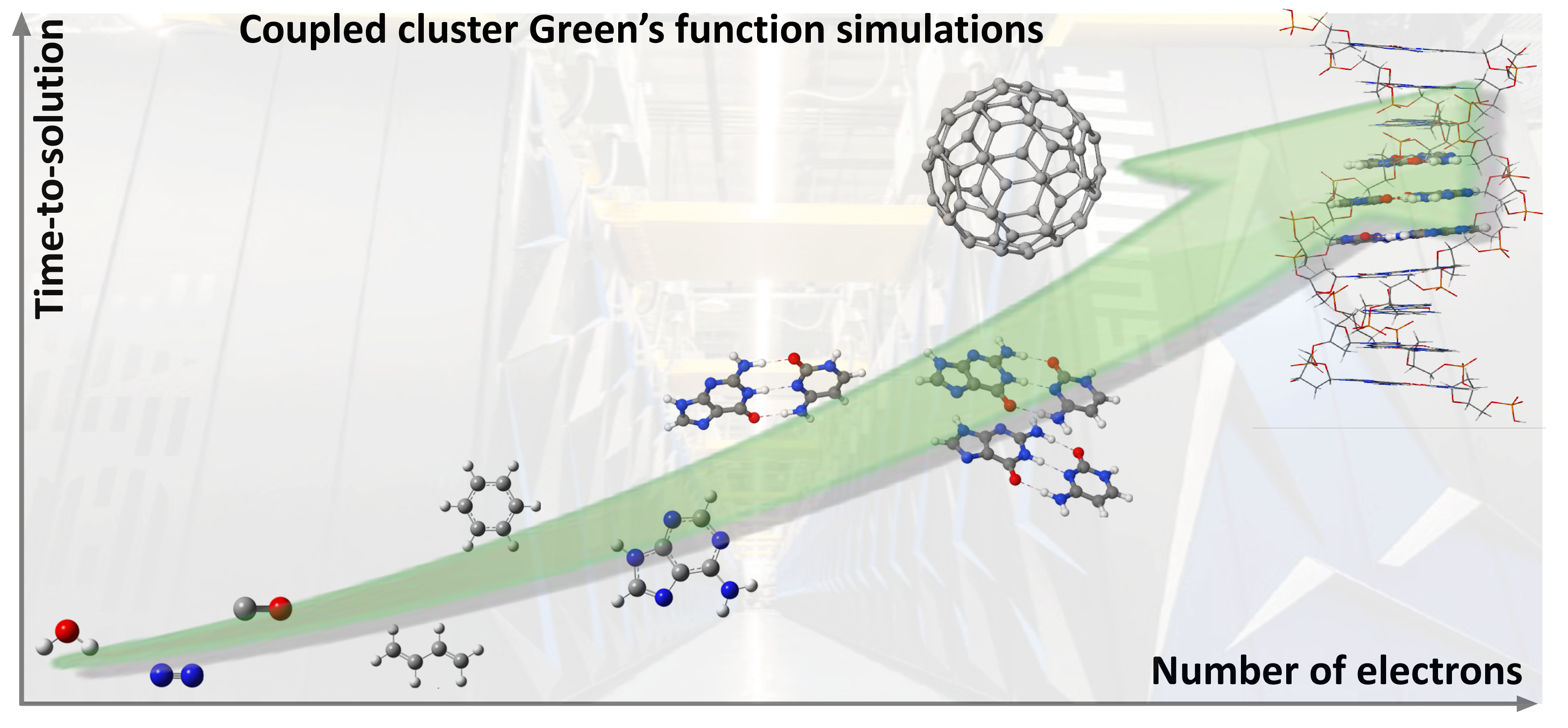}
    \caption{The CCGF approach facilitated by scalable implementation on the modern supercomputing facility is now developing towards large-scale simulations.}
    \label{CCSDGF_app}
\end{figure}

So far, the CCGF approach is mainly utilized for the accurate electronic structure studies of molecular systems and condensed-matter materials. Early implementation of CCGF formalism relying on the eigen-solvers was essentially carrying out ionization potential and electron attachment EOM-CC calculations to resolve the pole positions in the valence region. The CCSDGF results for predicting the ionization energies in the valence band of small benchmark systems such as Ne, HF, H$_2$O, N$_2$, and nucleobases employing triple-$\zeta$ basis sets agree well with the experimental photoelectron spectra data\cite{kowalski16_144101}. Further development of the CCGF approach involving linear-solvers enabled the computation of spectral functions for single atoms and small molecules over a broad frequency range. Employing the simple Gaussian basis set, the CCSDGF quasiparticle energy spectra of isolated atoms from H to Ne exhibited excellent agreement with the full-configuration interaction results reproducing not only correct quasiparticle peaks but also satellite peaks\cite{matsushita18_034106}. Similar calculations have also been carried out for 3$d$ transition metal atoms from V to Cu and simple periodic systems such as 1D LiH chain, 1D C chain, and 1D Be chain, and comparison has been made between CCSDGF  and GFs from single-particle theories such as Hartree-Fock and density functional theory (DFT)\cite{matsushita18_224103,matsushita18_204109} to exhibit the capability of the CCSDGF approach for computing the satellites of the spectral functions. The CCSDGF calculation of the ionization of H$_2$O, N$_2$, CO, s-trans-1,3-butadiene, benzene, and adenine molecules were found to be able to provide a qualitative or semi-quantitative level of description for the ionization processes in both the core and valence regimes\cite{kowalski18_4335}. Large-scale CCSDGF simulations were rarely reported until recently alongside the release of the scalable GFCCLib. In one of the reports\cite{peng20_011101}, the spectral functions of several guanine–cytosine (G–C) base pair structures have been reliably computed at the CCSD level employing the CCSDGF approach with 400$\sim$1200 basis functions for the first time in a relatively broad near-valence regime, where the many-body description of the near-valence ionization of the large DNA fragments features the important transition from the intra-base-pair cytosine $\pi\rightarrow\pi^\ast$ excitation to the inter-base-pair electron excitation as the fragment size increases. In another report\cite{peng2021GFCCLib}, CCSDGF calculations employing $>$800 basis functions were performed for the first time  for computing the many-body electronic structure of the fullerene C60 molecule covering up to $\sim$25 eV near-valence spectral region. It is worth mentioning that it has been observed from these reported large-scale CCSDGF simulations that good weak scaling behavior supports multiple GFCCSD calculations to be efficiently performed in parallel, while improved strong scaling can enable the use of fewer parallel tasks and thus reduce load imbalance.

For condensed-phase materials, an early test of the CCSDGF approach was performed for computing the spectral function of the uniform electron gas\cite{chan16_235139}, which was found to provide an improvement over GW and GW-cumulant results. More recently, the CCSDGF as an impurity solver has been applied in the embedding theoretical framework to compute the electronic structure of 1D and 2D Hubbard models\cite{zhu2019_115154,zgid19_6010} and limited solids\cite{zhu2021ab} exhibiting the reliability of CCSDGF solver in treating weakly and even moderately strongly correlated materials. For more realistic solids such as antiferromagnetic MnO$_2$ and paramagnetic SrMnO$_3$ whose insulating character the GW level of theory has difficulty predicting, recent examination of the corresponding impurity self-energies and local density-of-states has illustrated that the CCSDGF solver is able to provide a satisfactory description in particular for impurities containing Mn $t_{2g}$ and $e_g$ orbitals, while exhibit larger discrepancies for impurities containing O $2p$ orbitals concluding that higher-order excitations need to be included in the CCGF solver for describing stronger correlation\cite{yeh2021testing}. 

Regarding including higher-order excitations in CCGF calculations, take triples for example, due to the prohibitive $\mathcal{O}(N^8)$ scaling in the formal CCSDT/CCSDTGF approach, pragmatic choice could be focused on reduced cost iterative and non-iterative triples corrections. Remarkably, the triples corrections such as EOMIP-CCSD$^{\ast}$\cite{stanton99_8785,saeh1999application,manohar2009perturbative},
CCSDT-n (n=1,2,3)
\cite{bartlett85_4041,bartlett95_81,bartlett96_581,stanton99_8785,bomble2005equation}, 
CC3\cite{koch97_1808},  stochastic EOM-CC(P)\cite{Emiliano2019accurate,Yuwono2020accelerating}, EOM-CC(m,n)\cite{hirata00_255,hirata06_074111,krylov05_084107}, 
EOM-CCSD(T)(a)$^{\ast}$\cite{matthews2016new,jagau2018non}, and method-of-moments based non-iterative triple corrections\cite{kowalski2004new,Loch2006two}
have been extensively developed, and could be migrated to the CCGF framework. %
Recently, based on the analysis of the connectedness of the CCGF\cite{kowalski16_062512, kowalski18_561}, as well as the lack of the  size-intensivity in the EOM-CC(m,n) class of methods\cite{krylov05_084107}, Peng and Kowalski proposed to extend the excitation level of the inner auxiliary operators $X_p(\omega)$ and $Y_q(\omega)$ in the linear-solving CCSDGF approach to triples to implicitly include triples corrections to the CCSDGF results\cite{peng19_3185}. Preliminary results for computing the spectral functions of N$_2$ and CO molecules in the inner and outer valence regimes exhibited better agreement with the experimental results than the CCSDGF. So far, the benchmark studies of these triples corrections are still focused on small and simple molecular systems, the evaluation of their impact on the description of larger and/or strongly correlated systems, as well as what we can learn from the evaluation to come up with a general guideline for the applications, are still scientific and computational challenging, and might need to introduce further approximations. From this perspective, approximations such as active space~\cite{gour2005active,gour2006efficient,gour2006extension,ohtsuka2007active} and local-correlation/multi-level  approach~\cite{Li2009local,Li2010multilevel,Li2010improved} that have been studied in the EOM-CC framework can be directly tested and applied in the CCGF approach.

\section{Potential future applications}

During decades of development in theory, numerical approach, and software, we envision the applications of CCGF approach is reaching its prime. In the following, we highlight three areas in which one might witness CCGF applications in the near future.

\subsection{Quantum transport}

Recent years have witnessed the considerate progress in the experimental and theoretical studies of quantum transport in molecular nanostructures, in particular molecular junctions. To understand interesting phenomena observed in the experiments such as Coulomb blockade\cite{Lacroix_1981,Meir1991transport}, Franck-Condon blockade\cite{lecturcq2009electron,burzuri2014Franck}, spin blockade\cite{heersche2006electron}, Kondo effect\cite{Meir1993low}, quantum interference and decoherence\cite{ballmann2012experimental,vazquez2012probing,constant2012observation}, etc. the theoretical description is required to be formulated within the non-equilibrium many-body framework. This requirement becomes especially important when the qualitative physics incorporated in the Hamiltonian of the overall system cannot be properly described by an effective single-particle picture due to strong electron-electron and/or electron-vibronic interactions. On the other hand, the estimates of IPs and EAs for small molecules based on single-particle theory typically deviate from higher-level methods by 1 eV or more\cite{setten15_5665}. For larger molecules or metallic wires, the absolute error in IPs and EAs sometimes does not necessarily increase with the system size, particularly when the work function is dominated by a subsystem (e.g. a large metallic segment) and the DFT results are satisfactory. However, it is worth pointing out for higher computed IPs and EAs, as well as the alignments of these levels, the errors in these large molecular junctions remain large. Therefore, higher-level methods, such as the CCGF theory, should provide better and the next generation standard tools of ab initio transport calculations\cite{Evers2020Advances}. Here, developed scalable CCGF libraries (for example, GFCCLib) open prospects for treating strong electron-electron interaction and large extended molecular complex (including both molecule and parts of leads described by thousands of atoms and large enough basis sets) so that scalable simulations can be performed with size-converged computational parameters.

\subsection{Double ionization and double electron attachment}

Development of computationally robust and efficient wavefunction approaches which can handle open-shell and multi-reference situation remains one of the central topics in electronic-structure theory. Both double ionization potential (DIP) and double electron attachment (DEA) methods are economical for describing multi-reference systems. Nooijen and Bartlett introduced~\cite{dea_dip_marcel_bartlet} both approaches for similarity transformed EOM in 1997. Later on, numerous DIP/DEA-EOMCC economical development in high-order excitation manifold (e.g. with the design of active spaces), and their benchmark and applications have been reported~\cite{SATTELMEYER200342, marcel_dip_2008, monika_dea_2011, shen_dip_2013, shen2014doubly, ajala2017economical, stoneburner2017systematic, gulania_dea_2021,BOKHAN2018191}.
In DIP, the two electrons get removed from the $N$-electron reference state to obtain $N-2$ electron-deficient target state. On the other hand, in DEA, two electrons are attached to an $N$-electron reference to obtain $N+2$ electron target state. 

When two electrons distribute on top of a closed shell, forming open-shell species (also known as diradicals). The DEA method can describe the ground-state and excited-state of diradicals very efficiently. Spin-flip (SF) is another method that provides highly accurate singlet-triplet gaps of diradicals. However, for computing excitation energy of higher excited states in diradicals, DEA performs better compared to SF due to its extensive accessibility to fock-space~\cite{bond_interact_dea}.

The DIP method helps interpret Auger spectra. Although the intensities depend on the particular site of core ionization in Auger spectra, one can restrict the removal of one electron from the core and still get a good approximation of Auger peaks. Another application of DIP is the ionization spectra of doublet radical. First, ionization targets neutral radical and its excited states, while second, ionization targets the ionized radical and its excited states. One can also try targeting the multi-reference neutral by performing the DIP method on the dianionic reference~\cite{gulania_c2_dip_2019}. However, complications due to instability of the dianionic reference limit this path of studying the neutral~\cite{Tomasz_dip_stable_2011}. 

\begin{figure}[h!]
    \centering
    \includegraphics[scale=0.25]{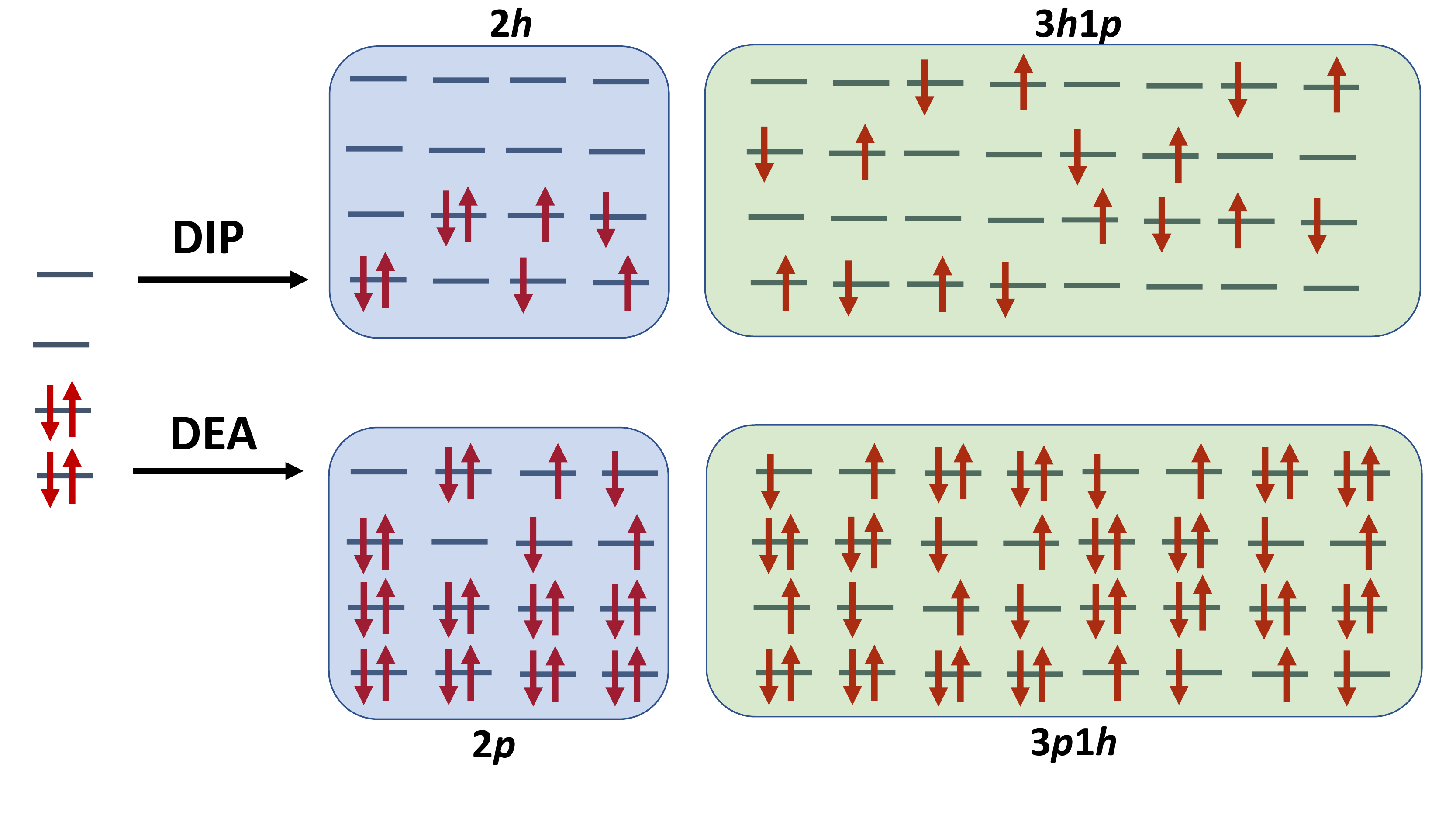}
    \caption{Singlet target state configurations generated by both DIP and DEA method from closed-shell reference}
    \label{fig:dip_dea_fock}
\end{figure}

Extending the capabilities of DEA and DIP within the CCGF approach will provide a new pathway of studying the electronic structure of multi-reference states spanning from low-energy to high energy regimes at one go. Furthermore, borrowing the ideas from EOMCCSD formalism and truncating the DEA and DIP operators at doubles excitation as follows:
\begin{equation}
    \hat{R}_{DIP} = \frac{1}{2}\sum_{ij}r_{ij}a_j a_i + \frac{1}{6}\sum_{ijka}r_{ijk}^{a} a_a^{\dagger} a_k a_j a_i
\end{equation}

\begin{equation}
    \hat{R}_{DEA} = \frac{1}{2}\sum_{ab}r^{ab}a_a^{\dagger}a_b^{\dagger} + \frac{1}{6}\sum_{iabc}r_{i}^{abc} a_a^{\dagger}a_b^{\dagger} a_c^{\dagger} a_i,
\end{equation}
where $r$ represents amplitude tensors, will provide tractable and easy-to-implement equations.

\subsection{Fingerprint spectroscopy}

\begin{figure}[h!]
    \includegraphics[width=\linewidth]{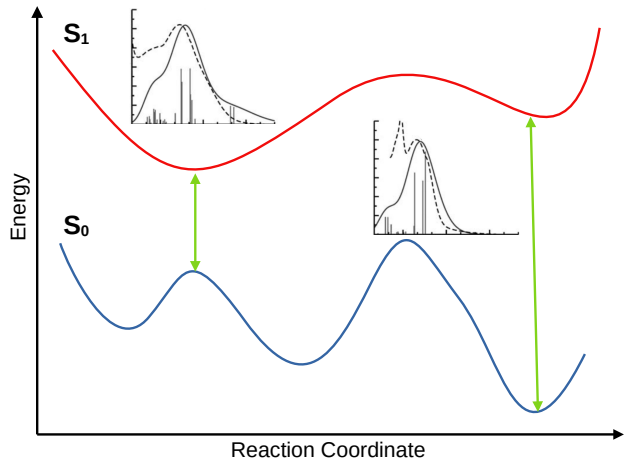}
    \caption{Abstract illustration of how CCGF methods can characterize fingerprint features of spectra.}
    \label{Fingerprint}
\end{figure}

The calculation of the vertical excitation spectrum, defined as the initial and final states having the same geometry (usually the equilibrium geometry of the ground state), is a natural starting point for investigating excited states of any polyatomic system. However, spectral features such as peak width, position, and intensity require more than interpretation at a single geometry. This is justifiable when we consider that electron ionization, attachment, and excitation process can result in potential energy landscapes that are very different from their parent surface, as abstractly illustrated in Fig. \ref{Fingerprint}\cite{adiabatic-NO3,adiabaticassess,adiabaticbases,BaumanAg,BaumanAu}. Techniques such as extended x-ray absorption fine structure, x-ray photoelectron diffraction, and core-level photoelectron spectroscopy are just some experimental means of analyzing structural and dynamical traits unique to molecules and characterizing regions of spectra known as fingerprint regions \cite{briggs1990practical,watts2019introduction,hufner2013photoelectron}. To accurately describe these distinct molecular spectroscopy motifs and their finer details, accurate computational methods need to be coupled with ways of analyzing the potential energy surface.

Analytical gradients based on CCGF approach have yet to be developed, but this does not preclude such methods from being combined with existing analytical gradient techniques of other approaches. For example, analytical gradients for various flavors of EOMCC approaches have been developed that can be leveraged. Most recently, a great deal of effort is being put forth to bring these analytical gradient techniques to utilize massively parallel architectures in the recent redesign of NWChem\cite{valiev2010nwchem} and NWChemEx\cite{nwchemex}. This implementation relies heavily on the many-body tensor library for the tensor contractions and the Libint library\cite{Libint} for the evaluation of molecular integrals over Gaussian Functions and their derivatives. Ideally, these standard methods of analyzing the potential energy surface, when combined with the CCGF methodology, ought to enhance the analysis and generation of spectra. 

\section{Conclusion}

In this work, following brief historical notes on the history of the development of Green's function method, in particular how it convoluted with wavefunction,  we reviewed early and recent development of CCGF theory and elaborated its theoretical foundation and some state-of-the-art numerical approaches, as well as its more recently scalable software development. At the end, we highlighted some of recent applications of CCGF approach and proposed some potential applications that would emerge in this field in the near future.

\section{Acknowledgement}
The work is supported by the Center for Scalable, Predictive methods for Excitation and Correlated phenomena (SPEC), which is funded by the U.S. Department of Energy (DOE), Office of Science, Office of Basic Energy Sciences, the Division of Chemical Sciences, Geosciences, and Biosciences. B. P. also acknowledges the support of the Laboratory Directed Research and Development (LDRD) program from PNNL. 

\bibliographystyle{ieeetr}
\bibliography{./gfcc}

\begin{thebibliography}{100}

\bibitem{green2008essay}
G.~Green, {\em An Essay on the Application of Mathematical Analysis to the
  Theories of Electricity and Magnetism}, p.~1–82.
\newblock Cambridge Library Collection - Mathematics, Cambridge University
  Press, 2014.

\bibitem{cannell2001Green}
D.~Cannell, {\em George Green: Mathematician and Physicist 1793–1841}.
\newblock Society for Industrial and Applied Mathematics, 2001.

\bibitem{Lindstrm2008OnTO}
J.~Lindstr{\"o}m, ``On the origin and early history of functional analysis,''
  2008.

\bibitem{grattan1995why}
I.~Grattan-Guinness, ``Why did george green write his essay of 1828 on
  electricity and magnetism?,'' {\em Am. Math. Mon.}, vol.~102, no.~5,
  pp.~387--396, 1995.

\bibitem{jahnkehistory}
H.~Jahnke, {\em A History of Analysis}.
\newblock History of mathematics, American Mathematical Soc., 2003.

\bibitem{nearing2010mathematical}
J.~Nearing, {\em Mathematical Tools for Physics}.
\newblock Dover books on mathematics, Dover Publications, 2010.

\bibitem{schrodinger2003collected}
E.~Schr{\"o}dinger, {\em Collected Papers on Wave Mechanics}.
\newblock AMS Chelsea Publishing Series, AMS Chelsea Pub., 2003.

\bibitem{amrein2005sturm}
W.~Amrein, A.~Hinz, and D.~Pearson, {\em Sturm-Liouville Theory: Past and
  Present}.
\newblock Birkh{\"a}user Basel, 2005.

\bibitem{dirac1933Lagrangian}
P.~Dirac, ``The lagrangian in quantum mechanics,'' {\em Phys. Z.}, vol.~3,
  pp.~64--72, 1933.

\bibitem{feynman1949theory}
R.~Feynman, ``The theory of positrons,'' {\em Phys. Rev.}, vol.~76,
  pp.~749--759, 1949.

\bibitem{feynman1958Space}
R.~Feynman, ``Space-time approach to non-relativistic quantum mechanics,'' {\em
  Rev. Mod. Phys.}, vol.~20, pp.~367--387, 1948.

\bibitem{schwinger1959Theory}
P.~Martin and J.~Schwinger, ``Theory of many-particle systems. i,'' {\em Phys.
  Rev.}, vol.~115, pp.~1342--1373, 1959.

\bibitem{baym1961conservation}
G.~Baym and L.~Kadanoff, ``Conservation laws and correlation functions,'' {\em
  Phys. Rev.}, vol.~124, pp.~287--299, 1961.

\bibitem{LB1}
R.~Landauer, ``Spatial variation of currents and fields due to localized
  scatterers in metallic conduction,'' {\em IBM J. Res. Develop.}, vol.~1,
  no.~3, pp.~223--231, 1957.

\bibitem{LB2}
R.~Landauer, ``Electrical resistance of disordered one-dimensional lattices,''
  {\em Phil. Mag.}, vol.~21, no.~172, pp.~863--867, 1970.

\bibitem{LB3}
M.~B\"uttiker, ``Absence of backscattering in the quantum hall effect in
  multiprobe conductors,'' {\em Phys. Rev. B}, vol.~38, pp.~9375--9389, 1988.

\bibitem{cizek66_4256}
J.~{\v C}{\'\i}{\v z}ek, ``On the correlation problem in atomic and molecular
  systems. calculation of wavefunction components in ursell-type expansion
  using quantum-field theoretical methods,'' {\em J. Chem. Phys.}, vol.~45,
  no.~11, pp.~4256--4266, 1966.

\bibitem{cizek1969}
J.~{\v C}{\'\i}{\v z}ek, ``On the use of the cluster expansion and the
  technique of diagrams in calculations of correlation effects in atoms and
  molecules,'' {\em Adv. Chem. Phys.}, vol.~2, pp.~35--89, 1969.

\bibitem{cizek1971}
J.~{\v C}{\'\i}{\v z}ek and J.~Paldus, ``Correlation problems in atomic and
  molecular systems iii. rederivation of the coupled-pair many-electron theory
  using the traditional quantum chemical methodst,'' {\em Int. J. Quantum
  Chem.}, vol.~5, no.~4, pp.~359--379, 1971.

\bibitem{paldus72_50}
J.~Paldus, J.~\ifmmode \check{C}\else \v{C}\fi{}\'{\i}\ifmmode~\check{z}\else
  \v{z}\fi{}ek, and I.~Shavitt, ``Correlation problems in atomic and molecular
  systems. iv. extended coupled-pair many-electron theory and its application
  to the b${\mathrm{h}}_{3}$ molecule,'' {\em Phys. Rev. A}, vol.~5,
  pp.~50--67, 1972.

\bibitem{nooijen92_55}
M.~Nooijen and J.~Snijders, ``Coupled cluster approach to the single-particle
  green's function,'' {\em Int. J. Quantum Chem.}, vol.~44, no.~S26,
  pp.~55--83, 1992.

\bibitem{nooijen93_15}
M.~Nooijen and J.~Snijders, ``Coupled cluster green's function method: Working
  equations and applications,'' {\em Int. J. Quantum Chem.}, vol.~48, no.~1,
  pp.~15--48, 1993.

\bibitem{nooijen95_1681}
M.~Nooijen and J.~Snijders, ``Second order many-body perturbation
  approximations to the coupled cluster green's function,'' {\em J. Chem.
  Phys.}, vol.~102, no.~4, pp.~1681--1688, 1995.

\bibitem{chan16_235139}
J.~McClain, J.~Lischner, T.~Watson, D.~Matthews, E.~Ronca, S.~Louie,
  T.~Berkelbach, and G.~Chan, ``Spectral functions of the uniform electron gas
  via coupled-cluster theory and comparison to the $gw$ and related
  approximations,'' {\em Phys. Rev. B}, vol.~93, p.~235139, 2016.

\bibitem{matsushita18_034106}
H.~Nishi, T.~Kosugi, Y.~Furukawa, and Y.~Matsushita, ``Quasiparticle energy
  spectra of isolated atoms from coupled-cluster singles and doubles (ccsd):
  Comparison with exact ci calculations,'' {\em J. Chem. Phys.}, vol.~149,
  no.~3, p.~034106, 2018.

\bibitem{matsushita18_224103}
T.~Kosugi, H.~Nishi, Y.~Furukawa, and Y.~Matsushita, ``Comparison of green's
  functions for transition metal atoms using self-energy functional theory and
  coupled-cluster singles and doubles (ccsd),'' {\em J. Chem. Phys.}, vol.~148,
  no.~22, p.~224103, 2018.

\bibitem{matsushita18_204109}
Y.~Furukawa, T.~Kosugi, H.~Nishi, and Y.~Matsushita, ``Band structures in
  coupled-cluster singles-and-doubles green's function (gfccsd),'' {\em J.
  Chem. Phys.}, vol.~148, no.~20, p.~204109, 2018.

\bibitem{kowalski18_4335}
B.~Peng and K.~Kowalski, ``Green's function coupled-cluster approach:
  Simulating photoelectron spectra for realistic molecular systems,'' {\em J.
  Chem. Theory Comput.}, vol.~14, no.~8, pp.~4335--4352, 2018.

\bibitem{berkelbach18_4224}
M.~Lange and T.~Berkelbach, ``On the relation between equation-of-motion
  coupled-cluster theory and the gw approximation,'' {\em J. Chem. Theory
  Comput.}, vol.~14, no.~8, pp.~4224--4236, 2018.

\bibitem{zhu2019_115154}
T.~Zhu, C.~Jim\'enez-Hoyos, J.~McClain, T.~Berkelbach, and G.~Chan,
  ``Coupled-cluster impurity solvers for dynamical mean-field theory,'' {\em
  Phys. Rev. B}, vol.~100, p.~115154, 2019.

\bibitem{zgid19_6010}
A.~Shee and D.~Zgid, ``Coupled cluster as an impurity solver for green’s
  function embedding methods,'' {\em J. Chem. Theory Comput.}, vol.~15, no.~11,
  pp.~6010--6024, 2019.

\bibitem{olcfsummit}
``Oak ridge leadership computing facility,'' 2020.

\bibitem{peng2021GFCCLib}
B.~Peng, A.~Panyala, K.~Kowalski, and S.~Krishnamoorthy, ``Gfcclib: Scalable
  and efficient coupled-cluster green's function library for accurately
  tackling many-body electronic structure problems,'' {\em Comput. Phys.
  Commun.}, vol.~265, p.~108000, 2021.

\bibitem{peng20_011101}
B.~Peng, K.~Kowalski, A.~Panyala, and S.~Krishnamoorthy, ``Green’s function
  coupled cluster simulation of the near-valence ionizations of
  dna-fragments,'' {\em J. Chem. Phys.}, vol.~152, no.~1, p.~011101, 2020.

\bibitem{meissner93_67}
L.~Meissner and R.~Bartlett, ``Electron propagator theory with the ground state
  correlated by the coupled-cluster method,'' {\em Int. J. Quantum Chem.},
  vol.~48, no.~S27, pp.~67--80, 1993.

\bibitem{kowalski14_094102}
K.~Kowalski, K.~Bhaskaran-Nair, and W.~A. Shelton, ``Coupled-cluster
  representation of green function employing modified spectral resolutions of
  similarity transformed hamiltonians,'' {\em J. Chem. Phys.}, vol.~141, no.~9,
  p.~094102, 2014.

\bibitem{kowalski16_144101}
K.~Bhaskaran-Nair, K.~Kowalski, and W.~Shelton, ``Coupled cluster green
  function: Model involving single and double excitations,'' {\em J. Chem.
  Phys.}, vol.~144, no.~14, p.~144101, 2016.

\bibitem{kowalski16_062512}
B.~Peng and K.~Kowalski, ``Coupled-cluster green's function: Analysis of
  properties originating in the exponential parametrization of the ground-state
  wave function,'' {\em Phys. Rev. A}, vol.~94, p.~062512, 2016.

\bibitem{kowalski18_561}
B.~Peng and K.~Kowalski, ``Properties of advanced coupled-cluster green's
  function,'' {\em Mol. Phys.}, vol.~116, no.~5-6, pp.~561--569, 2018.

\bibitem{kowalski18_214102}
B.~Peng and K.~Kowalski, ``Green's function coupled cluster formulations
  utilizing extended inner excitations,'' {\em J. Chem. Phys.}, vol.~149,
  no.~21, p.~214102, 2018.

\bibitem{keen2021hybrid}
T.~Keen, B.~Peng, K.~Kowalski, P.~Lougovski, and S.~Johnston, ``Hybrid
  quantum-classical approach for coupled-cluster green's function theory,''
  {\em arXiv preprint 2104.06981}, 2021.

\bibitem{rehr2020eom}
J.~Rehr, F.~Vila, J.~Kas, N.~Hirshberg, K.~Kowalski, and B.~Peng, ``Equation of
  motion coupled-cluster cumulant approach for intrinsic losses in x-ray
  spectra,'' {\em J. Chem. Phys.}, vol.~152, no.~17, p.~174113, 2020.

\bibitem{vila2020real}
F.~Vila, J.~Rehr, J.~Kas, K.~Kowalski, and B.~Peng, ``Real-time coupled-cluster
  approach for the cumulant green’s function,'' {\em J. Chem. Theory
  Comput.}, vol.~16, no.~11, pp.~6983--6992, 2020.

\bibitem{arno1951}
W.~Arnoldi, ``{The principle of minimized iterations in the solution of the
  matrix eigenvalue problem},'' {\em Quart. Appl. Math.}, vol.~9, pp.~17--29,
  1951.

\bibitem{bade2000}
Z.~Bai, J.~Demmel, J.~Dongarra, A.~Ruhe, and H.~{van der Vorst}, eds., {\em
  {Templates for the Solution of Algebraic Eigenvalue Problems: A Practical
  Guide}}.
\newblock Philadelphia, PA: Society for Industrial and Applied Mathematics,
  2000.

\bibitem{stanton95_1064}
J.~Stanton and J.~Gauss, ``Perturbative treatment of the similarity transformed
  hamiltonian in equation-of-motion coupled-cluster approximations,'' {\em J.
  Chem. Phys.}, vol.~103, no.~3, pp.~1064--1076, 1995.

\bibitem{cederbaum80_206}
L.~Cederbaum, W.~Domcke, and J.~Schirmer, ``Many-body theory of core holes,''
  {\em Phys. Rev. A}, vol.~22, pp.~206--222, 1980.

\bibitem{norman18_7208}
P.~Norman and A.~Dreuw, ``Simulating x-ray spectroscopies and calculating
  core-excited states of molecules,'' {\em Chem. Rev.}, vol.~118, no.~15,
  pp.~7208--7248, 2018.

\bibitem{schirmer87_6789}
G.~Angonoa, O.~Walter, and J.~Schirmer, ``Theoretical k-shell ionization
  spectra of n2 and co by a fourth-order green's function method,'' {\em J.
  Chem. Phys.}, vol.~87, no.~12, pp.~6789--6801, 1987.

\bibitem{dreuw14_4583}
J.~Wenzel, M.~Wormit, and A.~Dreuw, ``Calculating x-ray absorption spectra of
  open-shell molecules with the unrestricted algebraic-diagrammatic
  construction scheme for the polarization propagator,'' {\em J. Chem. Theory
  Comput.}, vol.~10, no.~10, pp.~4583--4598, 2014.

\bibitem{trofimov00_483}
A.~Trofimov, T.~Moskovskaya, E.~Gromov, N.~Vitkovskaya, and J.~Schirmer,
  ``Core-level electronic spectra in adc(2) approximation for polarization
  propagator: Carbon monoxide and nitrogen molecules,'' {\em J. Struct. Chem.},
  vol.~41, no.~3, pp.~483--494, 2000.

\bibitem{coriani12_1616}
S.~Coriani, T.~Fransson, O.~Christiansen, and P.~Norman,
  ``Asymmetric-lanczos-chain-driven implementation of electronic resonance
  convergent coupled-cluster linear response theory,'' {\em J. Chem. Theory
  Comput.}, vol.~8, no.~5, pp.~1616--1628, 2012.

\bibitem{peng15_4146}
B.~Peng, P.~Lestrange, J.~Goings, M.~Caricato, and X.~Li, ``Energy-specific
  equation-of-motion coupled-cluster methods for high-energy excited states:
  Application to k-edge x-ray absorption spectroscopy,'' {\em J. Chem. Theory
  Comput.}, vol.~11, no.~9, pp.~4146--4153, 2015.

\bibitem{krylov15_273}
D.~Zuev, E.~Vecharynski, C.~Yang, N.~Orms, and A.~Krylov, ``New algorithms for
  iterative matrix-free eigensolvers in quantum chemistry,'' {\em J. Comput.
  Chem.}, vol.~36, no.~5, pp.~273--284, 2015.

\bibitem{meyer73_1017}
W.~Meyer, ``Pno–ci studies of electron correlation effects. i. configuration
  expansion by means of nonorthogonal orbitals, and application to the ground
  state and ionized states of methane,'' {\em J. Chem. Phys.}, vol.~58, no.~3,
  pp.~1017--1035, 1973.

\bibitem{edmiston66_1833}
C.~Edmiston and M.~Krauss, ``Pseudonatural orbitals as a basis for the
  superposition of configurations. i. he2+,'' {\em J. Chem. Phys.}, vol.~45,
  no.~5, pp.~1833--1839, 1966.

\bibitem{edmiston68_192}
C.~Edmiston and M.~Krauss, ``Pseudonatural orbitals as a basis for the
  superposition of configurations. ii. energy surface for linear h3,'' {\em J.
  Chem. Phys.}, vol.~49, no.~1, pp.~192--205, 1968.

\bibitem{ahlrichs75_275}
R.~Ahlrichs and F.~Driessler, ``Direct determination of pair natural
  orbitals.,'' {\em Theor. Chim. Acta}, vol.~36, pp.~275--287, 1975.

\bibitem{neese13_034106}
C.~Riplinger and F.~Neese, ``An efficient and near linear scaling pair natural
  orbital based local coupled cluster method,'' {\em J. Chem. Phys.}, vol.~138,
  no.~3, p.~034106, 2013.

\bibitem{neese16_024109}
C.~Riplinger, P.~Pinski, U.~Becker, E.~Valeev, and F.~Neese, ``Sparse maps - a
  systematic infrastructure for reduced-scaling electronic structure methods.
  ii. linear scaling domain based pair natural orbital coupled cluster
  theory,'' {\em J. Chem. Phys.}, vol.~144, no.~2, p.~024109, 2016.

\bibitem{neese16_034102}
A.~Dutta, F.~Neese, and R.~Izs\'{a}k, ``Towards a pair natural orbital coupled
  cluster method for excited states,'' {\em J. Chem. Phys.}, vol.~145, no.~3,
  p.~034102, 2016.

\bibitem{neese18_244101}
A.~Dutta, M.~Saitow, C.~Riplinger, F.~Neese, and R.~Izs\'{a}k, ``A near-linear
  scaling equation of motion coupled cluster method for ionized states,'' {\em
  J. Chem. Phys.}, vol.~148, no.~24, p.~244101, 2018.

\bibitem{neese19_164123}
A.~Dutta, M.~Saitow, B.~Demoulin, F.~Neese, and R.~Izs\'{a}k, ``A domain-based
  local pair natural orbital implementation of the equation of motion coupled
  cluster method for electron attached states,'' {\em J. Chem. Phys.},
  vol.~150, no.~16, p.~164123, 2019.

\bibitem{yuichiro18}
T.~Kosugi and Y.~Matsushita, ``Wannier interpolation of one-particle green’s
  functions from coupled-cluster singles and doubles (ccsd),'' {\em J. Chem.
  Phys.}, vol.~150, no.~11, p.~114104, 2019.

\bibitem{peng19_3185}
B.~Peng, R.~van Beeumen, D.~Williams-Young, K.~Kowalski, and C.~Yang,
  ``Approximate green’s function coupled cluster method employing effective
  dimension reduction,'' {\em J. Chem. Theory Comput.}, vol.~15, no.~5,
  pp.~3185--3196, 2019.
\newblock PMID: 30951302.

\bibitem{pulay80_393}
P.~Pulay, ``Convergence acceleration of iterative sequences. the case of scf
  iteration,'' {\em Chem. Phys. Lett.}, vol.~73, no.~2, pp.~393--398, 1980.

\bibitem{pulay82_556}
P.~Pulay, ``Improved scf convergence acceleration,'' {\em J. Comput. Chem.},
  vol.~3, no.~4, pp.~556--560, 1982.

\bibitem{valiev2010nwchem}
M.~Valiev, E.~Bylaska, N.~Govind, K.~Kowalski, T.~Straatsma, H.~van Dam,
  D.~Wang, J.~Nieplocha, E.~Apr\'{a}, T.~Windus, and W.~de~Jong, ``Nwchem: A
  comprehensive and scalable open-source solution for large scale molecular
  simulations,'' {\em Comput. Phys. Commun.}, vol.~181, no.~9, pp.~1477--1489,
  2010.

\bibitem{nwchemex}
K.~Kowalski, R.~Bair, N.~Bauman, J.~Boschen, E.~Bylaska, J.~Daily, W.~de~Jong,
  T.~Dunning, N.~Govind, R.~Harrison, M.~Kebeli, K.~Keipert, S.~Krishnamoorthy,
  S.~Kumar, E.~Mutlu, B.~Palmer, A.~Panyala, B.~Peng, R.~Richard, T.~Straatsma,
  P.~Sushko, E.~Valeev, M.~Valiev, H.~van Dam, J.~Waldrop, D.~Williams-Young,
  C.~Yang, M.~Zalewski, and T.~Windus, ``From nwchem to nwchemex: Evolving with
  the computational chemistry landscape,'' {\em Chem. Rev.}, vol.~121, no.~8,
  pp.~4962--4998, 2021.

\bibitem{hirata039887}
S.~Hirata, ``Tensor contraction engine: Abstraction and automated parallel
  implementation of configuration-interaction, coupled-cluster, and many-body
  perturbation theories,'' {\em J. Phys. Chem. A}, vol.~107, no.~46,
  pp.~9887--9897, 2003.

\bibitem{hirata2006symbolic}
S.~Hirata, ``Symbolic algebra in quantum chemistry,'' {\em Theo. Chem. Acc.},
  vol.~116, no.~1-3, pp.~2--17, 2006.

\bibitem{deumens2011software}
E.~Deumens, V.~Lotrich, A.~Perera, M.~Ponton, B.~Sanders, and R.~Bartlett,
  ``Software design of aces iii with the super instruction architecture,'' {\em
  WIREs: Comput. Mol. Sci.}, vol.~1, no.~6, pp.~895--901, 2011.

\bibitem{deumens2011super}
E.~Deumens, V.~Lotrich, A.~Perera, R.~Bartlett, N.~Jindal, and B.~Sanders,
  ``The super instruction architecture: A framework for high-productivity
  parallel implementation of coupled-cluster methods on petascale computers,''
  {\em Annu. Rep. Comput. Chem.}, vol.~7, pp.~179--191, 2011.

\bibitem{solomonik2013cyclops}
E.~Solomonik, D.~Matthews, J.~Hammond, and J.~Demmel, ``Cyclops tensor
  framework: Reducing communication and eliminating load imbalance in massively
  parallel contractions,'' {\em 2013 IEEE 27th International Symposium on
  Parallel and Distributed Processing}, pp.~813--824, 2013.

\bibitem{solomonik2014massively}
E.~Solomonik, D.~Matthews, J.~Hammond, J.~Stanton, and J.~Demmel, ``A massively
  parallel tensor contraction framework for coupled-cluster computations,''
  {\em J. Parallel Distr. Comput.}, vol.~74, no.~12, pp.~3176--3190, 2014.

\bibitem{calvin2015scalable}
J.~Calvin, C.~Lewis, and E.~Valeev, ``Scalable task-based algorithm for
  multiplication of block-rank-sparse matrices,'' {\em Proceedings of the 5th
  Workshop on Irregular Applications: Architectures and Algorithms}, vol.~4,
  pp.~1--8, 2015.

\bibitem{peng2019coupled}
C.~Peng, J.~Calvin, and E.~Valeev, ``Coupled-cluster singles, doubles and
  perturbative triples with density fitting approximation for massively
  parallel heterogeneous platforms,'' {\em Int. J. Quantum Chem.}, vol.~119,
  no.~12, p.~e25894, 2019.

\bibitem{mutlu2019toward}
E.~Mutlu, K.~Kowalski, and S.~Krishnamoorthy, ``Toward generalized tensor
  algebra for ab initio quantum chemistry methods,'' in {\em Proceedings of the
  6th ACM SIGPLAN International Workshop on Libraries, Languages and Compilers
  for Array Programming}, pp.~46--56, 2019.

\bibitem{ga1994}
J.~Nieplocha, R.~Harrison, and R.~Littlefield, ``{Global Arrays: A Portable
  "Shared-memory" Programming Model for Distributed Memory Computers},'' in
  {\em Proceedings of the 1994 ACM/IEEE Conference on Supercomputing},
  Supercomputing 94, (Los Alamitos, CA, USA), pp.~340--349, IEEE Computer
  Society Press, 1994.

\bibitem{ga1995}
J.~Nieplocha, R.~Harrison, and R.~Littlefield, ``{The Global Array Programming
  Model for High Performance Scientific Computing},'' {\em SIAM News}, vol.~28,
  no.~7, pp.~12--14, 1995.

\bibitem{ga1996}
J.~Nieplocha, R.~Harrison, and R.~Littlefield, ``{Global arrays: A Nonuniform
  Memory Access Programming Model for High-Performance Computers},'' {\em J.
  Supercomput.}, vol.~10, no.~2, pp.~169--189, 1996.

\bibitem{TALSH}
D.~Lyakh, ``Talsh,'' 2014.

\bibitem{zhu2021ab}
T.~Zhu and G.~Chan, ``Ab initio full cell $gw+\mathrm{DMFT}$ for correlated
  materials,'' {\em Phys. Rev. X}, vol.~11, p.~021006, 2021.

\bibitem{yeh2021testing}
C.~Yeh, A.~Shee, S.~Iskakov, and D.~Zgid, ``Testing the green's function
  coupled cluster singles and doubles impurity solver on real materials within
  the framework of self-energy embedding theory,'' {\em Phys. Rev. B},
  vol.~103, p.~155158, 2021.

\bibitem{stanton99_8785}
J.~Stanton and J.~Gauss, ``A simple scheme for the direct calculation of
  ionization potentials with coupled-cluster theory that exploits established
  excitation energy methods,'' {\em J. Chem. Phys.}, vol.~111, no.~19,
  pp.~8785--8788, 1999.

\bibitem{saeh1999application}
J.~Saeh and J.~Stanton, ``Application of an equation-of-motion coupled cluster
  method including higher-order corrections to potential energy surfaces of
  radicals,'' {\em J. Chem. Phys.}, vol.~111, no.~18, pp.~8275--8285, 1999.

\bibitem{manohar2009perturbative}
P.~Manohar, J.~Stanton, and A.~Krylov, ``Perturbative triples correction for
  the equation-of-motion coupled-cluster wave functions with single and double
  substitutions for ionized states: Theory, implementation, and examples,''
  {\em J. Chem. Phys.}, vol.~131, no.~11, p.~114112, 2009.

\bibitem{bartlett85_4041}
M.~Urban, J.~Noga, S.~Cole, and R.~Bartlett, ``Towards a full ccsdt model for
  electron correlation,'' {\em J. Chem. Phys.}, vol.~83, no.~8, pp.~4041--4046,
  1985.

\bibitem{bartlett95_81}
J.~Watts and R.~Bartlett, ``Economical triple excitation equation-of-motion
  coupled-cluster methods for excitation energies,'' {\em Chem. Phys. Lett.},
  vol.~233, no.~1, pp.~81--87, 1995.

\bibitem{bartlett96_581}
J.~Watts and R.~Bartlett, ``Iterative and non-iterative triple excitation
  corrections in coupled-cluster methods for excited electronic states: the
  eom-ccsdt-3 and eom-ccsd(t) methods,'' {\em Chem. Phys. Lett.}, vol.~258,
  no.~5, pp.~581--588, 1996.

\bibitem{bomble2005equation}
Y.~Bomble, J.~Saeh, J.~Stanton, P.~Szalay, M.~K{\'a}llay, and J.~Gauss,
  ``Equation-of-motion coupled-cluster methods for ionized states with an
  approximate treatment of triple excitations,'' {\em J. Chem. Phys.},
  vol.~122, no.~15, p.~154107, 2005.

\bibitem{koch97_1808}
H.~Koch, O.~Christiansen, P.~J{\o}rgensen, A.~de~Merás, and T.~Helgaker, ``The
  cc3 model: An iterative coupled cluster approach including connected
  triples,'' {\em J. Chem. Phys.}, vol.~106, no.~5, pp.~1808--1818, 1997.

\bibitem{Emiliano2019accurate}
J.~E. Deustua, S.~H. Yuwono, J.~Shen, and P.~Piecuch, ``Accurate excited-state
  energetics by a combination of monte carlo sampling and equation-of-motion
  coupled-cluster computations,'' {\em J. Chem. Phys.}, vol.~150, no.~11,
  p.~111101, 2019.

\bibitem{Yuwono2020accelerating}
S.~H. Yuwono, A.~Chakraborty, J.~E. Deustua, J.~Shen, and P.~Piecuch,
  ``Accelerating convergence of equation-of-motion coupled-cluster computations
  using the semi-stochastic cc(p;q) formalism,'' {\em Mol. Phys.}, vol.~118,
  no.~19-20, p.~e1817592, 2020.

\bibitem{hirata00_255}
S.~Hirata, M.~Nooijen, and R.~Bartlett, ``High-order determinantal
  equation-of-motion coupled-cluster calculations for electronic excited
  states,'' {\em Chem. Phys. Lett.}, vol.~326, no.~3, pp.~255--262, 2000.

\bibitem{hirata06_074111}
M.~Kamiya and S.~Hirata, ``Higher-order equation-of-motion coupled-cluster
  methods for ionization processes,'' {\em J. Chem. Phys.}, vol.~125, no.~7,
  p.~074111, 2006.

\bibitem{krylov05_084107}
L.~Slipchenko and A.~Krylov, ``Spin-conserving and spin-flipping
  equation-of-motion coupled-cluster method with triple excitations,'' {\em J.
  Chem. Phys.}, vol.~123, no.~8, p.~084107, 2005.

\bibitem{matthews2016new}
D.~Matthews and J.~Stanton, ``A new approach to approximate equation-of-motion
  coupled cluster with triple excitations,'' {\em J. Chem. Phys.}, vol.~145,
  no.~12, p.~124102, 2016.

\bibitem{jagau2018non}
T.~Jagau, ``Non-iterative triple excitations in equation-of-motion
  coupled-cluster theory for electron attachment with applications to bound and
  temporary anions,'' {\em J. Chem. Phys.}, vol.~148, no.~2, p.~024104, 2018.

\bibitem{kowalski2004new}
K.~Kowalski and P.~Piecuch, ``New coupled-cluster methods with singles,
  doubles, and noniterative triples for high accuracy calculations of excited
  electronic states,'' {\em J. Chem. Phys.}, vol.~120, no.~4, pp.~1715--1738,
  2004.

\bibitem{Loch2006two}
M.~W. Łoch, M.~D. Lodriguito, P.~Piecuch, and J.~R. Gour, ``Two new classes of
  non-iterative coupled-cluster methods derived from the method of moments of
  coupled-cluster equations,'' {\em Mol. Phys.}, vol.~104, no.~13-14,
  pp.~2149--2172, 2006.

\bibitem{gour2005active}
J.~R. Gour, P.~Piecuch, and M.~Włoch, ``Active-space equation-of-motion
  coupled-cluster methods for excited states of radicals and other open-shell
  systems: Ea-eomccsdt and ip-eomccsdt,'' {\em J. Chem. Phys.}, vol.~123,
  no.~13, p.~134113, 2005.

\bibitem{gour2006efficient}
J.~R. Gour and P.~Piecuch, ``Efficient formulation and computer implementation
  of the active-space electron-attached and ionized equation-of-motion
  coupled-cluster methods,'' {\em J. Chem. Phys.}, vol.~125, no.~23, p.~234107,
  2006.

\bibitem{gour2006extension}
J.~R. Gour, P.~Piecuch, and M.~Włoch, ``Extension of the active-space
  equation-of-motion coupled-cluster methods to radical systems: The
  ea-eomccsdt and ip-eomccsdt approaches,'' {\em Int. J. Quantum Chem.},
  vol.~106, no.~14, pp.~2854--2874, 2006.

\bibitem{ohtsuka2007active}
Y.~Ohtsuka, P.~Piecuch, J.~R. Gour, M.~Ehara, and H.~Nakatsuji, ``Active-space
  symmetry-adapted-cluster configuration-interaction and equation-of-motion
  coupled-cluster methods for high accuracy calculations of potential energy
  surfaces of radicals,'' {\em J. Chem. Phys.}, vol.~126, no.~16, p.~164111,
  2007.

\bibitem{Li2009local}
W.~Li, P.~Piecuch, J.~R. Gour, and S.~Li, ``Local correlation calculations
  using standard and renormalized coupled-cluster approaches,'' {\em J. Chem.
  Phys.}, vol.~131, no.~11, p.~114109, 2009.

\bibitem{Li2010multilevel}
W.~Li and P.~Piecuch, ``Multilevel extension of the cluster-in-molecule local
  correlation methodology: Merging coupled-cluster and møller-plesset
  perturbation theories,'' {\em J. Phys. Chem. A}, vol.~114, no.~24,
  pp.~6721--6727, 2010.

\bibitem{Li2010improved}
W.~Li and P.~Piecuch, ``Improved design of orbital domains within the
  cluster-in-molecule local correlation framework: Single-environment
  cluster-in-molecule ansatz and its application to local coupled-cluster
  approach with singles and doubles,'' {\em J. Phys. Chem. A}, vol.~114,
  no.~33, pp.~8644--8657, 2010.

\bibitem{Lacroix_1981}
C.~Lacroix, ``Density of states for the anderson model,'' {\em J. Phys. F:
  Metal Phys.}, vol.~11, no.~11, pp.~2389--2397, 1981.

\bibitem{Meir1991transport}
Y.~Meir, N.~Wingreen, and P.~Lee, ``Transport through a strongly interacting
  electron system: Theory of periodic conductance oscillations,'' {\em Phys.
  Rev. Lett.}, vol.~66, pp.~3048--3051, 1991.

\bibitem{lecturcq2009electron}
R.~Leturcq, C.~Stampfer, K.~Inderbitzin, L.~Durrer, C.~Hierold, E.~Mariani,
  M.~Schultz, F.~von Oppen, and K.~Ensslin, ``Franck–condon blockade in
  suspended carbon nanotube quantum dots,'' {\em Nature Phys.}, vol.~5,
  pp.~327--331, 2009.

\bibitem{burzuri2014Franck}
E.~Burzurí, Y.~Yamamoto, M.~Warnock, X.~Zhong, K.~Park, A.~Cornia, and
  H.~van~der Zant, ``Franck–condon blockade in a single-molecule
  transistor,'' {\em Nano Lett.}, vol.~14, no.~6, pp.~3191--3196, 2014.

\bibitem{heersche2006electron}
H.~B. Heersche, Z.~de~Groot, J.~A. Folk, H.~S.~J. van~der Zant, C.~Romeike,
  M.~R. Wegewijs, L.~Zobbi, D.~Barreca, E.~Tondello, and A.~Cornia, ``Electron
  transport through single ${\mathrm{mn}}_{12}$ molecular magnets,'' {\em Phys.
  Rev. Lett.}, vol.~96, p.~206801, 2006.

\bibitem{Meir1993low}
Y.~Meir, N.~Wingreen, and P.~Lee, ``Low-temperature transport through a quantum
  dot: The anderson model out of equilibrium,'' {\em Phys. Rev. Lett.},
  vol.~70, pp.~2601--2604, 1993.

\bibitem{ballmann2012experimental}
S.~Ballmann, R.~H\"artle, P.~Coto, M.~Elbing, M.~Mayor, M.~Bryce, M.~Thoss, and
  H.~Weber, ``Experimental evidence for quantum interference and vibrationally
  induced decoherence in single-molecule junctions,'' {\em Phys. Rev. Lett.},
  vol.~109, p.~056801, 2012.

\bibitem{vazquez2012probing}
H.~Vazquez, R.~Skouta, S.~Schneebeli, M.~Kamenetska, R.~Breslow,
  L.~Venkataraman, and M.~Hybertsen, ``Probing the conductance superposition
  law in single-molecule circuits with parallel paths,'' {\em Nature
  Nanotech.}, vol.~7, pp.~663--667, 2012.

\bibitem{constant2012observation}
C.~Gu\'{e}don, H.~Valkenier, T.~Markussen, K.~Thygesen, J.~Hummelen, and
  S.~van~der Molen, ``Observation of quantum interference in molecular charge
  transport,'' {\em Nature Nanotech..}, vol.~7, pp.~305--309, 2012.

\bibitem{setten15_5665}
M.~van Setten, F.~Caruso, S.~Sharifzadeh, X.~Ren, M.~Scheffler, F.~Liu,
  J.~Lischner, L.~Lin, J.~Deslippe, S.~Louie, C.~Yang, F.~Weigend, J.~Neaton,
  F.~Evers, and P.~Rinke, ``Gw100: Benchmarking g0w0 for molecular systems,''
  {\em J. Chem. Theory Comput.}, vol.~11, no.~12, pp.~5665--5687, 2015.

\bibitem{Evers2020Advances}
F.~Evers, R.~Koryt\'ar, S.~Tewari, and J.~van Ruitenbeek, ``Advances and
  challenges in single-molecule electron transport,'' {\em Rev. Mod. Phys.},
  vol.~92, p.~035001, 2020.

\bibitem{dea_dip_marcel_bartlet}
M.~Nooijen and R.~Bartlett, ``Similarity transformed equation-of-motion
  coupled-cluster theory: Details, examples, and comparisons,'' {\em J. Chem.
  Phys.}, vol.~107, no.~17, pp.~6812--6830, 1997.

\bibitem{SATTELMEYER200342}
K.~Sattelmeyer, H.~{Schaefer III}, and J.~Stanton, ``Use of 2h and 3h-p-like
  coupled-cluster tamm-dancoff approaches for the equilibrium properties of
  ozone,'' {\em Chem. Phys. Lett.}, vol.~378, no.~1, pp.~42--46, 2003.

\bibitem{marcel_dip_2008}
O.~Demel, K.~Shamasundar, L.~Kong, and M.~Nooijen, ``Application of double
  ionization state-specific equation of motion coupled cluster method to
  organic diradicals,'' {\em J. Phys. Chem. A}, vol.~112, no.~46,
  pp.~11895--11902, 2008.

\bibitem{monika_dea_2011}
M.~Musiał, S.~Kucharski, and R.~Bartlett, ``Multireference double electron
  attached coupled cluster method with full inclusion of the connected triple
  excitations: Mr-da-ccsdt,'' {\em J. Chem. Theory Comput.}, vol.~7, no.~10,
  pp.~3088--3096, 2011.

\bibitem{shen_dip_2013}
J.~Shen and P.~Piecuch, ``Doubly electron-attached and doubly ionized
  equation-of-motion coupled-cluster methods with 4-particle–2-hole and
  4-hole–2-particle excitations and their active-space extensions,'' {\em J.
  Chem. Phys.}, vol.~138, no.~19, p.~194102, 2013.

\bibitem{shen2014doubly}
J.~Shen and P.~Piecuch, ``Doubly electron-attached and doubly ionised
  equation-of-motion coupled-cluster methods with full and active-space
  treatments of 4-particle–2-hole and 4-hole–2-particle excitations: the
  role of orbital choices,'' {\em Mol. Phys.}, vol.~112, no.~5-6,
  pp.~868--885, 2014.

\bibitem{ajala2017economical}
A.~O. Ajala, J.~Shen, and P.~Piecuch, ``Economical doubly electron-attached
  equation-of-motion coupled-cluster methods with an active-space treatment of
  three-particle–one-hole and four-particle–two-hole excitations,'' {\em J.
  Phys. Chem. A}, vol.~121, no.~18, pp.~3469--3485, 2017.

\bibitem{stoneburner2017systematic}
S.~J. Stoneburner, J.~Shen, A.~O. Ajala, P.~Piecuch, D.~G. Truhlar, and
  L.~Gagliardi, ``Systematic design of active spaces for multi-reference
  calculations of singlet–triplet gaps of organic diradicals, with benchmarks
  against doubly electron-attached coupled-cluster data,'' {\em J. Chem.
  Phys.}, vol.~147, no.~16, p.~164120, 2017.

\bibitem{gulania_dea_2021}
S.~Gulania, E.~Kjønstad, J.~Stanton, H.~Koch, and A.~Krylov,
  ``Equation-of-motion coupled-cluster method with double electron-attaching
  operators: Theory, implementation, and benchmarks,'' {\em J. Chem. Phys.},
  vol.~154, no.~11, p.~114115, 2021.

\bibitem{BOKHAN2018191}
D.~Bokhan, D.~Trubnikov, A.~Perera, and R.~Bartlett, ``Explicitly-correlated
  double ionization potentials and double electron attachment
  equation-of-motion coupled cluster methods,'' {\em Chem. Phys. Lett.},
  vol.~692, pp.~191--195, 2018.

\bibitem{bond_interact_dea}
M.~Ivanov, S.~Gulania, and A.~Krylov, ``Two cycling centers in one molecule:
  Communication by through-bond interactions and entanglement of the unpaired
  electrons,'' {\em J. Phys. Chem. Lett.}, vol.~11, no.~4, pp.~1297--1304,
  2020.

\bibitem{gulania_c2_dip_2019}
S.~Gulania, T.~Jagau, and A.~Krylov, ``{EOM-CC} guide to fock-space travel: The
  {C}$_2$ edition,'' {\em Faraday Discuss.}, vol.~217, pp.~514--532, 2019.

\bibitem{Tomasz_dip_stable_2011}
T.~Kus and A.~Krylov, ``Using the charge-stabilization technique in the double
  ionization potential equation-of-motion calculations with dianion
  references,'' {\em J. Chem. Phys.}, vol.~135, no.~8, p.~084109, 2011.

\bibitem{adiabatic-NO3}
W.~Eisfeld and K.~Morokuma, ``Ab initio investigation of the vertical and
  adiabatic excitation spectrum of no3,'' {\em J. Chem. Phys.}, vol.~114,
  no.~21, pp.~9430--9440, 2001.

\bibitem{adiabaticassess}
R.~Send, M.~Kühn, and F.~Furche, ``Assessing excited state methods by
  adiabatic excitation energies,'' {\em J. Chem. Theory Comput.}, vol.~7,
  no.~8, pp.~2376--2386, 2011.

\bibitem{adiabaticbases}
K.~Bravaya, O.~Kostko, S.~Dolgikh, A.~Landau, M.~Ahmed, and A.~Krylov,
  ``Electronic structure and spectroscopy of nucleic acid bases: Ionization
  energies, ionization-induced structural changes, and photoelectron spectra,''
  {\em J. Phys. Chem. A}, vol.~114, no.~46, pp.~12305--12317, 2010.

\bibitem{BaumanAg}
N.~Bauman, J.~Hansen, and P.~Piecuch, ``Coupled-cluster interpretation of the
  photoelectron spectrum of ag3-,'' {\em J. Chem. Phys.}, vol.~145, no.~8,
  p.~084306, 2016.

\bibitem{BaumanAu}
N.~Bauman, J.~Hansen, M.~Ehara, and P.~Piecuch, ``Coupled-cluster
  interpretation of the photoelectron spectrum of au3-,'' {\em J. Chem. Phys.},
  vol.~141, no.~10, p.~101102, 2014.

\bibitem{briggs1990practical}
D.~Briggs and M.~Seah, {\em Practical Surface Analysis, Auger and X-ray
  Photoelectron Spectroscopy}.
\newblock Practical Surface Analysis, John Wiley \& Sons, Ltd, 1990.

\bibitem{watts2019introduction}
J.~F. Watts and J.~Wolstenholme.
\newblock John Wiley \& Sons, Ltd, 2019.

\bibitem{hufner2013photoelectron}
S.~H{\"u}fner, {\em Photoelectron Spectroscopy: Principles and Applications}.
\newblock Springer Series in Solid-State Sciences, Springer Berlin Heidelberg,
  2013.

\bibitem{Libint}
E.~Valeev, ``Libint: A library for the evaluation of molecular integrals of
  many-body operators over gaussian functions.'' http://libint.valeyev.net/.
\newblock [Online].

\end{thebibliography}

\end{document}